\documentclass[preprints,article,accept,pdftex,moreauthors]{mdpi}

\firstpage{1}
\pubvolume{1}
\issuenum{1}
\articlenumber{0}
\pubyear{2022}
\copyrightyear{2022}
\externaleditor{Academic Editor: Sergio Cristallo}
\datereceived{10 July 2022}
\daterevised{26 August 2022} 
\dateaccepted{28 August 2022}
\datepublished{}

\hreflink{https://doi.org/10.3390/\linebreak universe1833788}
\doinum{10.3390/universe1833788}

\pdfoutput=1

\Title{The C/M Ratio of AGB Stars in the Local Group Galaxies}

\TitleCitation{The C/M Ratio of AGB Stars in the Local Group Galaxies}

\Author{{Tongtian Ren} $^{1,2}$\orcidA{}, Biwei Jiang $^{1,2,}$*\orcidB{}, Yi Ren $^{3}$\orcidC{} and Ming Yang $^{4,5}$\orcidD{}} 

\AuthorNames{Tongtian Ren, Biwei Jiang, Yi Ren and Ming Yang}

\AuthorCitation{Ren,T.; Jiang, B.; Ren, Y.; Yang, M.}

\address{%
$^{1}$ \quad Institute for Frontiers in Astronomy and Astrophysics, Beijing Normal University, Beijing 102206, China; rentongtian@mail.bnu.edu.cn\\
$^{2}$ \quad Department of Astronomy, Beijing Normal University, Beijing 100875, China\\
$^{3}$ \quad College of Physics and Electronic Engineering, Qilu Normal University, Jinan 250200, China; yiren@qlnu.edu.cn\\
$^{4}$ \quad Key Laboratory of Space Astronomy and Technology, National Astronomical Observatories, Chinese Academy of Sciences, Beijing 100101, China; myang@nao.cas.cn\\
$^{5}$ \quad IAASARS, National Observatory of Athens, Vas. Pavlou and I. Metaxa, Penteli 15236, Greece}

\corres{Correspondence: bjiang@bnu.edu.cn}

\abstract{The number ratio of carbon-rich to oxygen-rich asymptotic giant branch (AGB) stars (the so-called C/M ratio) is closely related to the evolution environment of the host galaxy. This work studies the C/M ratio in 14 galaxies within the Local Group with the most complete and clean sample of member stars identified in our previous works. The borderlines between carbon-rich AGB and oxygen-rich AGB stars as well as red supergiants are defined by Gaussian mixture model fitting to the number density in the $(J-K)/K$ diagram for the member stars of the LMC and M33, and  then applied to the other galaxies by shifting the difference in the position of tip red giant branch (TRGB).  The C/M ratios are obtained  after precise and consistent categorization. Although for galaxies with larger distance modulo there is greater uncertainty, the C/M ratio is clearly found to decrease with the color index $(J-K)_0$ of TRGB as the indicator of metallicity, which agrees with previous studies and can be explained by the fact that carbon stars are more easily formed in a metal-poor environment. Furthermore, the C/M ratio within M33 is found to increase with galactocentric distance, which coincides with this scenario and the galactic chemical evolution model. On the other hand, the C/M ratio within M31 is found to decrease with galactocentric radius, which deserves further study. }


\begin{document}

\defcitealias{ren2021red}{Paper I}
\defcitealias{ren2021rsglocal}{Paper II}

\section{Introduction} \label{section1}

Asymptotic Giant Branch (AGB) stars are the low- to intermediate-mass \mbox{($\sim$0.8 to 8 ${M_\odot}$) } stars in the late evolutionary stage \citep{herwig2005evolution,pinte2009benchmark}. During this stage, the stars own electron-degenerated and inert carbon-oxygen cores, which are surrounded by shells of He and H burning alternatively \citep{hofner2018mass}, while huge envelopes can extend to 1 AU in their outer layer. According to the Stefan–Boltzmann law, such huge sizes result in 
 the high luminosities of the AGB stars, whose traces in the Hertzsprung-Russell Diagram (HRD) are almost aligned with {the Red Giant Branch (RGB) stars with hydrogen fusing in shells. Evidently, the AGB stars are generally brighter than the tip of the RGB stars (TRGB), where the RGB stars reach their brightness upper limit} \citep{vassiliadis1993evolution}.

The AGB phase can be divided into two stages: the early AGB (E-AGB) stage and thermally pulsing AGB (TP-AGB) stage \citep{iben1983asymptotic}. During the TP-AGB phase, {the third dredge-up (3DU) episodes bring the products of the He-burning nucleosynthesis, including carbon as well as the elements heavier than Fe synthesized by the slow neutron capture process, into the stellar surface}, 
changing the abundance of stellar atmospheres \citep{busso1999nucleosynthesis,fishlock2014evolution,karakas2014dawes,cristallo2015evolution}. Initially, the atmospheres of AGB stars are dominated by oxygen, but the 3DU process in the TP-AGB stage would bring out a gradual increase in the abundance of carbon \citep{weiss2009new}. Based on {the relative abundance of carbon and oxygen (C/O), AGB stars can be classified into C-rich AGB (C-AGB hereafter, C/O$>$1) stars and O-rich AGB (O-AGB hereafter, C/O$<$1) stars. In addition, the S-type AGB stars between the two types have strong s-element enrichments \citep{trippella2014s,trippella2016s,busso2021s}, and their C/O ratio can be as low as 0.5 \citep{smolders2012spitzer}. Their near-infrared colors are similar to that of O-AGB stars \citep{cioni2003agb}, so in this work we count them as O-AGB stars. In O-AGB stars, the strong absorption features of oxygen-bearing molecules such as TiO and H$_{2}$O are present,} while optical spectra of C-AGB stars are characterized by CH, CN, SiC$_{2}$ or other carbon-bearing molecules. {These unique spectral features make it possible to distinguish C- and O-AGB stars by spectroscopy, and Father Angelo Secchi first discovered C-AGB stars by such features in the 1860s} 
\citep{secchi1868catalogue}.

In terms of the spectral type, C-AGB stars are of N- or C-type, while O-AGB stars are of M- or K-type \citep{cioni2003agb}. Therefore, the number ratio of C-AGB and O-AGB stars is expressed as C/M ratio. Theoretically, the C/M ratio of a group of stars is influenced by stellar metallicity, age and mass. {It is easier for an O-rich star to become C-rich in a low metallicity environment} \citep{battinelli2005calibration,cioni2005near}. {For low-mass AGB stars, the 3DU is not so efficient, while for massive stars, the Hot Bottom Burning (HBB) prevents their transition to the C-rich stage \citep{iben1983asymptotic}. From the compilation of \citet{frogel1990asymptotic}, it can be seen that the luminous C-AGB stars are absent in the oldest clusters of the Magellanic Clouds (MCs). Besides, they are also
uncommonly found in the young metal-rich population. This suggests the existence of lower and upper mass limits of C-AGBs, respectively \citep{marigo2017new}.}

In observation, there are two conventional methods to distinguish C-AGBs and O-AGBs, i.e. the narrow-band filter method and the color-magnitude diagram method. The narrow-band filter method makes use of the spectral features around the TiO-7780{\AA} band and the CN-8120{\AA} band of O-AGBs and C-AGBs, respectively.  The color index [TiO]-[CN] in the narrow-bands filters centered at 7780 and 8120{\AA} then reveals the type of stars being C-rich or O-rich. \citet{palmer1982new} successfully applied this method to 47 Tuc by stellar positions in the [TiO]-[CN] vs $V-I$ diagram  to distinguish C-AGBs and O-AGBs. Subsequently, this method is applied in several studies (e.g., \citep{cook1986carbon,brewer1995late,brewer1996late,letarte2002extent}). However, the TiO and CN filters are not popular equipment for telescopes, which limits the generalization of this method. Both [TiO] and [CN] are located at the wavelengths of optical bands, where the brightness of AGB stars is fainter than in the infrared bands, which is negative for the detection of the spectral features. Moreover, narrow filters require longer exposure time to collect enough photons. The above disadvantages limit the completeness of the narrow-bands method. This method is also susceptible to contamination by radiation from foreground M dwarfs and background galaxies \citep{letarte2002extent,groenewegen2006agb,demers2006carbon}. The CN spectral lines may get covered by other absorption lines when the C stars are reddened by surrounding dust containing myriad molecules, resulting in an underestimated number of C-AGB stars.

An alternative method is based on the near-infrared color-magnitude diagrams (CMD), because the C-AGBs are apparently redder and brighter than the O-AGBs. With the near-infrared CMD method, the stellar content in many nearby galaxies were investigated (e.g.,~\citep{davidge1998evolved,davidge2000evolved,davidge2003asymptotic,davidge2004photometric,davidge2005disk}). \citet{cioni2003agb,cioni2005near,cioni2008agb} have classified the AGB stars in the Magellanic Clouds, M33, NGC 6822 and derived their C/M ratios with the data of the infrared (IR) surveys such as DENIS or 2MASS.  In more recent studies, \citet{sibbons2012agb,sibbons2015agb} calculated the C/M ratios of NGC 6822 and IC 1613 also via this method.

A problem {that} cannot be ignored with the near-infrared CMD method is that O-rich AGB stars are vulnerable to contamination by foreground M dwarfs, which damages the purity of the AGB sample and may lead to over-estimating O-AGB stars and then underestimating the C/M ratio. Besides, these studies usually set some constant color indexes as the criteria for the borderline between O-AGBs, C-AGBs, and x-AGBs (extreme AGB with thick circumstellar envelop). However, the distribution of various types of stars in the CMD is additionally influenced by the metallicity of the host galaxy \citep{battinelli2007assessment}, thus making it inappropriate to apply the same partition borders for different galaxies. In fact, a skewed distribution matches  more closely the reality. \citet{pastorelli2019constraining,pastorelli2020constraining} calibrated the AGB population in the Magellanic Clouds based on 2MASS and GAIA photometric measurements, their samples clearly reveal a near-parallel sloped distribution on CMDs. The $J-K/K$ diagram of the complete sample of red supergiant (RSG) stars and AGB stars in M31 and M33 also show that RSG stars and AGB stars present skewed distributions in the CMD in \citet{ren2021red}.

This work intends to {study systematically} the C/M ratio of fourteen galaxies in the Local Group and re-calibrate the relation of this ratio with metallicity. The Local Group contains a few dozens of member galaxies, and new faint members are still being discovered, which span a wide range of morphology, star formation history and metallicity. These galaxies' relatively short distances (mostly within 1Mpc) make it possible to resolve and survey stellar populations in different evolutionary stages, in particular the luminous giants such as red supergiants (RSGs) and AGB stars. Thus, the characteristics of evolutionary patterns of stars in galaxies of various environments can be revealed. Based on the mediocrity principle, the stellar sample of galaxies in the Local Group allows us to make reasonable assumptions about the stellar populations of further distant galaxies for which stars are hard to be resolved.

{The major improvements of this work lie in two aspects: the new samples of AGB stars in the Local Group Galaxies and the method to classify the AGB stars. In \citet{ren2021red} (Paper I hereafter) and \citet{ren2021rsglocal} (Paper II hereafter), we have established the most complete and pure sample of AGB stars in the targeted galaxies after removing the foreground dwarf stars effectively with a novel method based on the UKIRT near-infrared color-color diagram. Furthermore, we designed a semi-automatic code based on the Gaussian mixture model to categorize the RSG and various types of AGB stars more accurately and reliably. }

The paper is structured as follows. First the data are described and selected in \mbox{Section \ref{section2}}, the method of classification is outlined in Section \ref{section3}, the result and discussion is presented in Section \ref{section4}, and the summary is given in Section \ref{section5}.

\section{Data: The Sample of AGB Stars}  \label{section2}

This study is based on the sample of red giants and supergiants including {RGBs}, AGBs and RSGs in 14 Local Group galaxies after removing the foreground dwarfs in \citetalias{ren2021red} and \citetalias{ren2021rsglocal}. The host galaxies cover various distances and types, including two irregular satellite galaxies closest to the Milky Way (the LMC and the SMC), two large spiral galaxies (M31 and M33), five dwarf spheroidal galaxies (NGC 147, NGC 185, Pegasus Dwarf, Sextans A and Sextans B), and five dwarf irregular galaxies (IC 10, IC 1613, NGC 6822, WLM and Leo A).

{The dwarf stars have higher surface gravity and higher density than giant stars, so the molecules in their atmospheres can form more easily \citep{allard1996model}. As a result, the molecules form at higher temperatures in dwarfs, causing absorption in the H band and the $H$ band brightness decreasing. Eventually, dwarfs have smaller $J-H$ and bigger $H-K$ than giants, a clear bifurcation is visible on the $J-H/H-K$ diagram \citep{bessell1988jhklm}. In this preliminary sample of \citetalias{ren2021red} and \citetalias{ren2021rsglocal}, the foreground dwarf stars are identified by taking use of this bifurcation.} In addition, the proper motion and parallax measurements by Gaia/EDR3 are supplemented to remove dwarfs or to identify the member stars depending on the distance of the galaxy. Furthermore, the interstellar extinction is corrected for the member stars. Because these galaxies span a large range of type and Galactic latitude, \citetalias{ren2021rsglocal} applied various methods for extinction correction depending on the individual galaxy, including the reddening map by SFD98 \citep{schlegel1998maps}, Bayestar19 \citep{green20193d} and OGLE \citep{skowron2021ogle}. {However, no extinction correction is applied for the sample stars of M31 and M33, because their extinction maps suffer relatively large uncertainty and lead to confusion instead of clarification on the CMD. In addition, their extinction is small (see Section \ref{section34} for more)}.

Accordingly, we divide the sample into three groups: one group includes only the MCs, one group includes the ten dwarf galaxies at a much further distance, and another group includes M31 and M33 -- two large spiral galaxies. Table \ref{table1} presents the $K_0$ magnitude, the color index $(J-K)_0$ of {TRGB} after the extinction correction, and the number of member stars of each galaxy in the initial sample from \mbox{\citetalias{ren2021red}and} \citetalias{ren2021rsglocal}. In addition, the [Fe/H] of each galaxy is derived from the TRGB location according to the calibration between the colors and [M/H] of the RGB stars \citep{bellazzini2004calibration}, and listed in Table \ref{table1}.

\begin{table}[H]
\caption{The basic information of the 14 sample galaxies. \label{table1}}
	\begin{adjustwidth}{-\extralength}{0cm}
		\newcolumntype{C}{>{\centering\arraybackslash}X}
		\begin{tabularx}{\fulllength}{CCCCCCCC}
			\toprule
		\textbf{Galaxy} &{$\textbf{\emph{K}}_{\textbf{0}}^\textbf{TRGB}$} & ${\textbf{\emph{(J}}-\textbf{\emph{K)}}}_{\textbf{0}}^\textbf{TRGB}$ & \textbf{[M/H] \textsuperscript{1}} &\textbf{Number of Sources (Initial Sample)} &$\sigma_{\textbf{\emph{JHK}}}$ & \textbf{Number of Sources (Final Sample)} & \textbf{{\textmu}}\\
			\midrule
			LMC & 12.00 & 1.00 & $-$1.00 & 402,300 & 0.05 & 198,548 & 18.26\\
			SMC & 12.71 & 0.91 & $-$1.24 & 66,573 & 0.05 & 29,413 & 18.72\\
			NGC 6822 & 17.38 & 0.98 & $-$1.05 & 4,084 & 0.20 & 3,745 & 23.58\\
			NGC 185 & 17.62 & 1.02 & $-$0.95 & 8,015 & 0.20 & 6,267 & 23.94\\
			IC10   & 18.14 & 0.86 & $-$1.37 & 15,336 & 0.20 & 12,368 & 24.01\\
			NGC 147 & 17.87 & 1.03 & $-$0.92 & 10,567 & 0.20 & 8,393 & 24.21\\
			IC 1613 & 18.10 & 0.96 & $-$1.11 & 2,363 & 0.20 & 1,970 & 24.25\\
			M31 & 17.66 & 1.12 & $-$0.68 & 1,245,930 & 0.20 & 513,957 & 24.25\\
			Leo A   & 18.56 & 0.90 & $-$1.26 & 86 & 0.20 & 72 & 24.54\\
			M33 & 18.17 & 1.07 & $-$0.82 & 203,486 & 0.20 & 95,812 & 24.62\\
			WLM     & 18.68 & 0.94 & $-$1.16 & 890 & 0.20 & 685 & 24.77\\
			Pegasus DIG & 18.65 & 1.02 & $-$0.95 & 398 & 0.20 & 297 & 24.97\\
			Sextans A & 20.01 & 0.80 & $-$1.53 & 194 & 0.20 & 173 & 25.95\\
			Sextans B & 20.21 & 0.80 & $-$1.53 & 141 & 0.20 & 97 & 26.15\\
			\bottomrule
		\end{tabularx}
	\end{adjustwidth}
	\noindent{\footnotesize{\textsuperscript{1} The [M/H] of each galaxy is derived from the ${(J-K)}_{0}^\mathrm{TRGB}$ - [M/H] relation in \citet{bellazzini2004calibration}.}}
\end{table}

\subsection{The Magellanic Clouds} \label{section21}

With a distance much shorter than the other twelve galaxies, the LMC and the SMC locate at 49.59 $\pm$ 0.09 (stat.) $\pm$ 0.54 (syst.) kpc \citep{pietrzynski2019distance} and 62.44 kpc $\pm$ 0.47 (stat.) $\pm$ 0.81 (syst.) kpc \citep{graczyk2020distance}, respectively. They are the two closest galaxies in our sample and their $JHK$ magnitudes are taken from the Two Micron All-Sky Survey (2MASS) \citep{skrutskie2006two}. For the convenience of comparison with other galaxies, the 2MASS $JHK_s$ magnitudes are converted to the UKIRT system according to the transformation equations in \citet{hodgkin2009ukirt}. From the LMC and SMC catalogues of \citetalias{ren2021rsglocal}, the following criteria are applied to select the sample to study the C/M ratio: (1) the identified member stars in \citetalias{ren2021rsglocal}, and (2) the photometric accuracy is limited to better than 0.05 mag in all the $JHK$ bands. Because AGB and RSG stars in the Magellanic Cloud are brighter than 12.0 and 12.7 mag respectively, i.e., much brighter than the sensitivity limit of 2MASS, this accuracy limit has little effect on the C/M ratio. On the other hand, the strict constraint on photometry ensures the precise division between various types of red giants.

\subsection{Ten Dwarf Galaxies} \label{section22}

The ten dwarf galaxies are all much more distant than the Magellanic Clouds with a distance modulus mostly from 24 to 25 mag so the 2MASS survey was not sensitive enough to reach the TRGB for most of them. Instead, their near-infrared $JHK$ photometry data come from the UKIRT/WFCAM observation with limiting magnitude at about $K=19$ mag which is much deeper than the 2MASS limit. Correspondingly, the criteria to select the stars are modified as: (1) the identified member stars in \citetalias{ren2021rsglocal}; and (2) the photometric accuracy is better than 0.2 mag in all the ${JHK}$ bands; (3) the extended sources are removed. We add the new filter: N\_Flag = 3, or N\_Flag =  2 \endnote{The N\_flag is added in \citetalias{ren2021rsglocal} to indicate the number of bands in which the source is identified as a point source.} , and it is not an extended source. \citetalias{ren2021rsglocal} missed this criterion, which caused the overestimation of the RSG stars' numbers. It will be shown by the fewer number of RSG stars compared to {\citetalias{ren2021rsglocal}} later in Section \ref{section43}. 

\subsection{M31 and M33} \label{section23}

M31 and M33 are two large spiral galaxies with distances beyond the detection limit of 2MASS. Their near-infrared $JHK$ photometries are also taken from the UKIRT/WFCAM. The criteria to select the stars from them are: (1) the member stars identified in \citetalias{ren2021red}, and located in the ellipse area defined by B=25 mag/arcsec$^2$ isophotes; and (2) the photometric accuracy is better than 0.2 mag in all the ${JHK}$ bands.

Table \ref{table1} presents the number of sources in the preliminary and final samples, respectively.

\section{Classification of Evolved Stars}  \label{section3}

\subsection{Selection of the Benchmark Galaxies: The LMC and M33} \label{section31}

The type of evolved stars can be defined by the location in the $(J-K)/K$ CMD that serves as an HR diagram, and the NIR bands are sensitive to the evolved stars. \citetalias{ren2021red} combine the observational and theoretical features in this CMD to clearly classify the evolved stars in M31 and M33 based on eye-check. The basic rule is to take the position of TRGB as the starting point, where RSGs are brighter and bluer than TRGB while AGB stars are brighter and redder than TRGB. We follow this rule in principle, and make some fine adjustments according to the specific situation.

With the selected sources in Section \ref{section2}, the $(J-K)/K$ CMD is created to analyze the type of stars. For the dwarf galaxies, the number of sources is generally too few to display distinct features in this CMD. Meanwhile, for large galaxies such as M31, there is neither clear CMD contour for different types of stars, which is caused by relatively confusing photometry in the crowded fields. In order to solve such problems, we select two benchmark galaxies which own enough stars with reliable photometry to define the borderlines between different types of stars. The borderlines from the benchmark galaxies are then applied to other galaxies with the shifts calculated from the TRGB position.

As stated before, the LMC has numerous stars to reveal the locations in the $(J-K)_0/K_0$ CMD of various types of evolved stars such as RGBs, RSGs, O-AGBs, C-AGBs and x-AGBs. In Section \ref{section2}, the  12 low-mass galaxies are divided into two groups by distance, that is the MCs, and the other ten distant ones, which depend on the photometric sensitivity and uncertainty of 2MASS and UKIRT/WFCAM, respectively. Here we continue to use this division because the photometric error would broaden some branches in the CMD and thus affect the determination of the borderline. On the other hand, M31 and M33 are large galaxies with many more stars and their CMDs become more heterogeneous than other galaxies. {Besides, unlike the dwarf galaxies, the extinctions of them are not corrected before we categorize their sources. Thus we set M31 and M33 into a new group for which M33 is selected as a benchmark galaxy because of its relatively clear CMD contour (Figure \ref{fig2}) and lower extinction \citep{wang2021red}.}

\subsection{Definition of the Borderlines for the LMC}\label{section32}

Because each type of stars cluster in some area of the $(J-K)/K$ CMD, the borderlines are usually located at the minimum number density of the diagram. The method to find the minimum density differs and depends on the quality of data. Some authors define the borderlines by eye-check, e.g., \citetalias{ren2021red}, while some define it by density histogram analysis e.g., \citet{cioni2005near,sibbons2012agb}. However, the true borders between different components on the CMD are sloped, some sophisticated methods are developed, e.g., \citet{hirschauer2020dusty} introduced the kernel density estimate (KDE) technique. In a case irrelevant to the C/M AGB analysis, \citet{rosenfield2016evolution} used a dual Gaussian fitting to select TP-AGB stars, which is more reliable with a single minimum at the trough of dual Gaussian curves which corresponds to the bound of two types of stars. {We take the above two methods for reference, and use the Gaussian mixture model (GMM) for classification. GMM is a statistical modeling method fitted by the maximum likelihood estimate using the Expectation-maximization (EM) algorithm, which decomposes a multi-peaked probability distribution into a linear combination of several single Gaussian distributions \citep{fraley2002model}. We then use a semi-automatic technology based on GMM, by classifying the types of stars in each $K$ magnitude cut on the CMD by which we  obtain the borderlines by linear fitting the boundaries in each bin.}

With a short distance to the Milky Way, the LMC has the most complete and best-measured sample of evolved stars and it is commonly selected as the calibrator of stellar models \citep{pastorelli2019constraining,pastorelli2020constraining}. There are 198,548 sources with $\sigma_{JHK}\le$ 0.05 mag in the final sample. The~$(J-K)_0/K_0$ diagram of these sources in  Figure \ref{fig1} clearly illustrates various types of evolved stars above TRGB: RSGs brighter and bluer than TRGB, O-AGBs brighter and redder than TRGB, C-AGBs even redder than O-AGBs that extend out like a tail \citep{marigo2003red} at $K \approx$ 11 mag, and the most luminous sources are RSGs and TP-AGB stars ($K_0 <$ 10 mag). In~accordance with the obvious branches of RSGs and various types of AGB stars, the $K$ band magnitude is firstly divided into nine cuts in the $K_0$ brightness whose center positions are 11.8, 11.7, 11.0, 10.7, 10.0, 9.5, 9.2, 8.5 and 8.0. The two faintest cuts ($K_0=11.7$, $K_0=11.8$) divide RSG and AGB stars at the faint position, while the two brightest cuts ($K_0=8.0$, $K_0=8.5$) constrain the location of the brightest RSG stars. Moreover, four relatively faint cuts ($K_0=9.5$, $K_0=10.0$, $K_0=10.7$,$K_0=11.0$) are crucial to dividing C-AGBs and O-AGBs, and a relatively luminous cut ($K_0=9.2$) is effective for discriminating RSG stars and TP-AGB stars. Regarding the width of each cut, it varies. Though a large bin width is needed to include more sources for reliable statistical analysis in the low-density region, a too-large bin width would trigger an inaccurate determination of the minimum position due to the sloped distribution of the stars on the CMD.

As Figure \ref{fig1} shows, the number of sources in the bin can vary by an order of magnitude. A test bin width of $\pm$ 0.05 mag is run first, yielding the number of sources in each bin $N_{\mathrm{Test}}$. The adopted bin width is then obtained by the following calculation:
\begin{equation}
    \mathrm{binwidth} = 1/\left( \ln N_\mathrm{Test}\right)
\end{equation}

\begin{figure}[H]
\begin{adjustwidth}{-\extralength}{0cm}
    \centering
    \includegraphics[width=14cm]{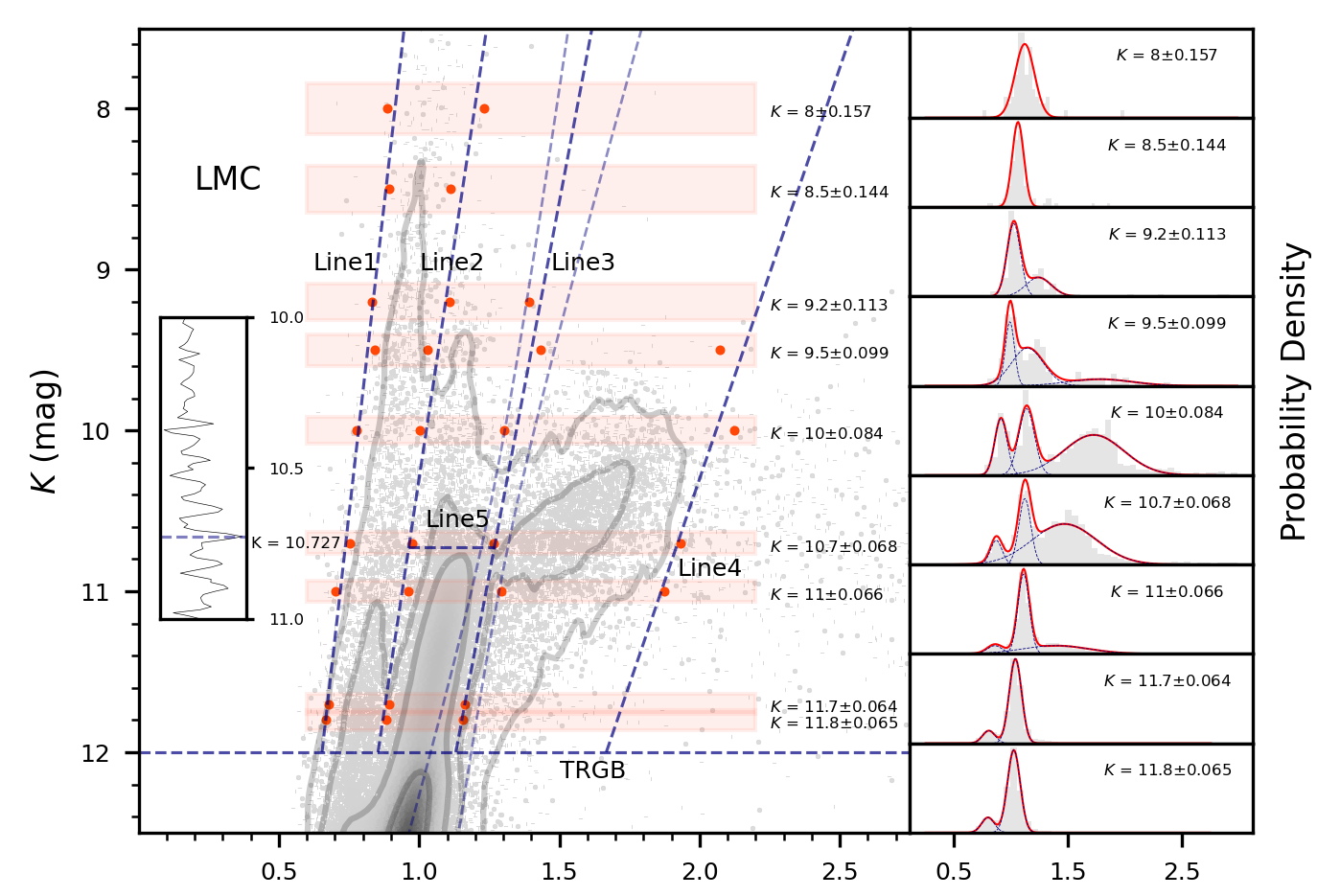}
    \end{adjustwidth}
    \caption{\textit{Cont}.   \label{fig1}}
\end{figure}

\begin{figure}[H]\ContinuedFloat
\begin{adjustwidth}{-\extralength}{0cm}
    \centering
    \includegraphics[width=14cm]{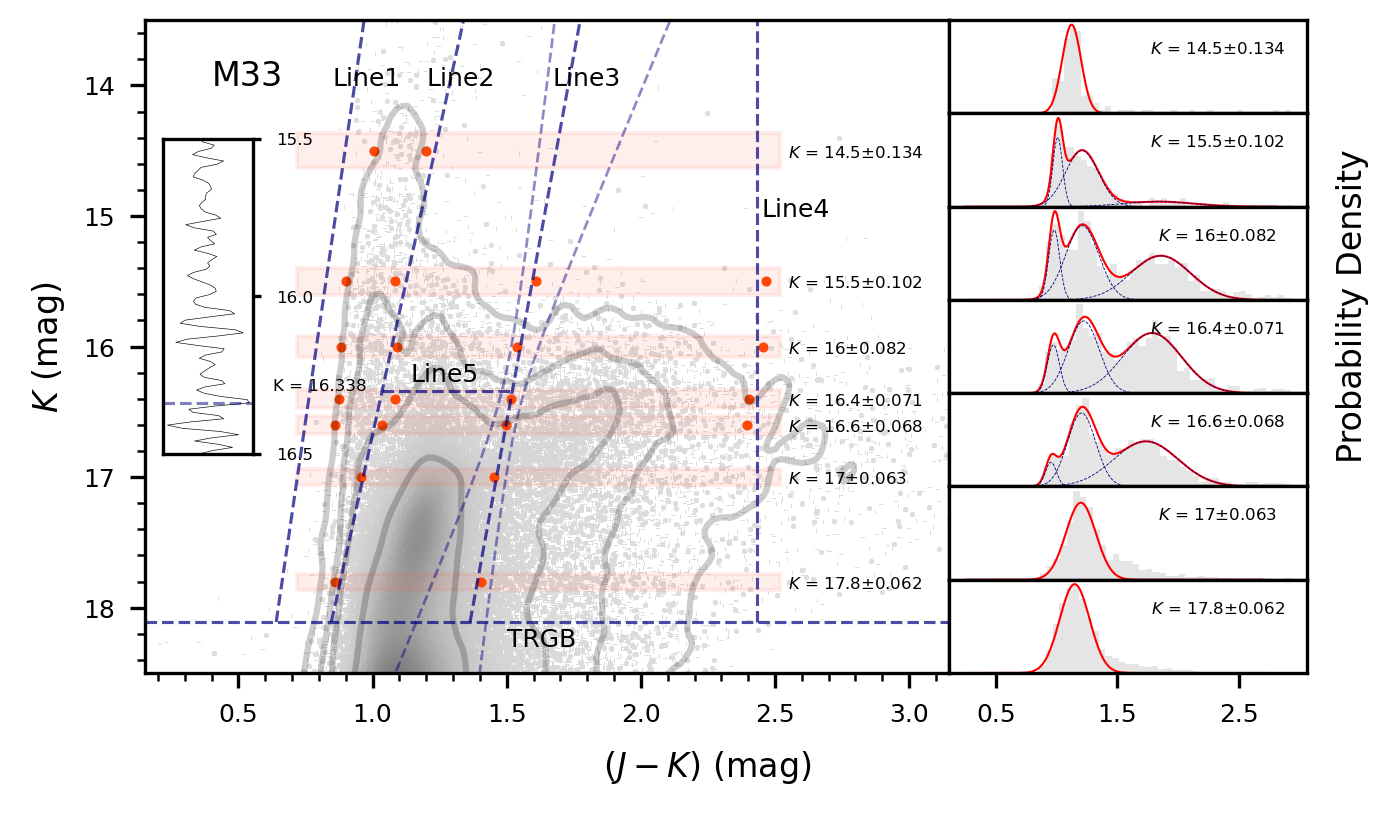}
        \end{adjustwidth}
\caption{Determination of the borderlines between different types of stars in the $(J-K)_0/K_0$ diagram in the LMC and M33. The boundaries (red dots) of each magnitude cut (red masks) calculated from GMM fitting (right panels), are used to fit the borderlines (dashed line). The dashed lines on either side of Line 3 represent the location of its 50\% confidence interval. The insert panels show the reference for detecting the $K$ magnitudes borderlines (Line 5) between O-AGBs and TP-AGBs with the Sobel filter.
    \label{fig1}}
\end{figure}

All resultant bin widths are smaller than 0.2 mag as displayed in Figure \ref{fig1}, and the subsequent distinctions between the Gaussian peaks are guaranteed to be found even for the magnitude cut with fewer than 100 sources. Then we detect the number of components and borders within each magnitude cuts based on GMM. In practice, the EM algorithm of GMM is stochastic in its application, and the output estimated parameters are not constant or may even deviate apparently from the true distribution. In addition, the number of components {from GMM} is easily overestimated, and only up to three components in each magnitude cut is allowed in this work in accordance with the true number of types of stars. In addition, this algorithm is unsuitable for determining the borderline (Line 5) between the TP-AGB tail and the fainter O-AGBs.

Basically, we use the software sklearn.mixture.GaussianMixture \endnote{\url{https://scikit-learn.org/stable/modules/generated/sklearn.mixture.GaussianMixture.html} (accessed on March 23, 2022)  
} with some modification to analyze the boundaries of different components. The initial number of components is all set as three, and then the model is run 1000 times to obtain the weights, centers, and covariances of each component, as well as the average and standard deviation of each parameter. Afterwards, the model is run 1000 times again, but we discard the output parameters beyond the standard deviation calculated by the above process, which left a group of stable and reliable triple Gaussian distribution results.

For the average of the three parameters of weights, mean locations, and covariances, we discarded the components that had no peak characteristics in the real data. Only two types of components are selected: (1) with weights over 90\%; (2) components with weights larger than 5\% and less than 90\%, and covariances less than 0.09 (i.e., widths less than 0.3) because they represent the peaks for certain components. As shown in Figure \ref{fig1}, based on the above semi-automatic selection, the number of components within each cut on the CMD of the LMC is obtained, as indicated by the number of peaks in the distribution curve on the right sides.

The troughs between each peak represent the boundaries between the different components. The hundreds of results of each run of the {GMM} are random in nature, so the $(J-K)_0$ color indices corresponding to the troughs of the output results also vary. We calculate the mean and standard deviation of the troughs and their uncertainties. Within each magnitude cut, the uncertainty of all troughs $(J-K)_0$ is found below 0.01mag.

On the edge, the blue borderline of RSGs and the red borderline of C-AGBs need to be determined individually. The solution is to decompose the histogram at  $K_0=$ 9.5, 10.0, 10.7 and 11.0 mag into three Gaussian components in which the leftmost and rightmost ones represent the distributions of RSG stars and C-AGBs, respectively. The $(J-K)_0$ color index that corresponds to the percent point function (ppf) = 0.01 for the leftmost Gaussian component (RSG) and ppf = 0.95 for the rightmost Gaussian component (C-AGBs) is taken as the blue and the red limit, respectively. For the dual Gaussian components at $K_0=$ 9.2, 11.7 and 11.8, the histogram is decomposed into two Gaussian functions representing  RSG stars and O-AGBs, and the ppf = 0.01 $\&$ 0.99 positions are taken as the blue and red limit, respectively. The two brightest cuts at $K_0=8.0$ and $8.5$ include only one Gaussian component from RSGs, and the criteria of ppf = 0.1 $\&$ 0.9 are used for the determination of its $(J-K)_0$ boundaries due to their larger dispersion.

With the trough points and the blue or red limits determined in each bin, the borderlines (Line 1, Line 2, Line 3, Line 4) are obtained by linear fitting to distinguish different types of stars. The Sobel filter is commonly used to detect the edge of the luminosity function, especially the brightness of {TRGB} \citep{da1990standard,lee1993tip}. We apply it to {detect} the brightness border between TP-AGB stars and O-AGBs. We smoothed the number distribution of the stars between Line 2 and Line 3 with a Gaussian kernel function of width 0.01, and then detected the edge at $K_0 = 10.727$ with the Sobel filter. This defined brightness is used as the borderline between TP-AGB and O-AGBs, i.e., Line 5.

However, it cannot be ignored that the uncertainties of the borderlines will affect the result of the C/M ratio. The most influential  borderline is Line 3 between O- and C-AGBs, while other borderlines are less significant. With the orthogonal distance regression (ODR) algorithm, we obtain the potential location area of Line 3 within the 50\% confidence interval (Figure \ref{fig1}), which also introduces the asymmetric error in the C/M ratio (Table \ref{table3}). It should be noted that the uncertainty of the blueward shift of Line 3 may lead to a large number of O-AGBs being identified as C-AGBs, increasing the derived C/M ratio considerably. In reverse, with the shifting of Line 3 towards the redder color, the decreasing in C-AGBs and the increasing in O-AGB are less, because this change in the classification of the targeted sources occurs in the low-density region. Therefore, the decreasing in the C/M ratio is lower than the increasing due to the uncertainty, resulting in asymmetric errors (Table \ref{table3}). This asymmetry is more evident in the SMC than in the LMC. In M33, another benchmark galaxy, the uncertainty is particularly obvious at the faint end (see Section \ref{section34}).

\subsection{The Borderlines of the SMC and Ten Dwarf Galaxies}\label{section33}

\begin{figure}[H]
\begin{adjustwidth}{-\extralength}{0cm}
    \centering
    \includegraphics[width=16cm]{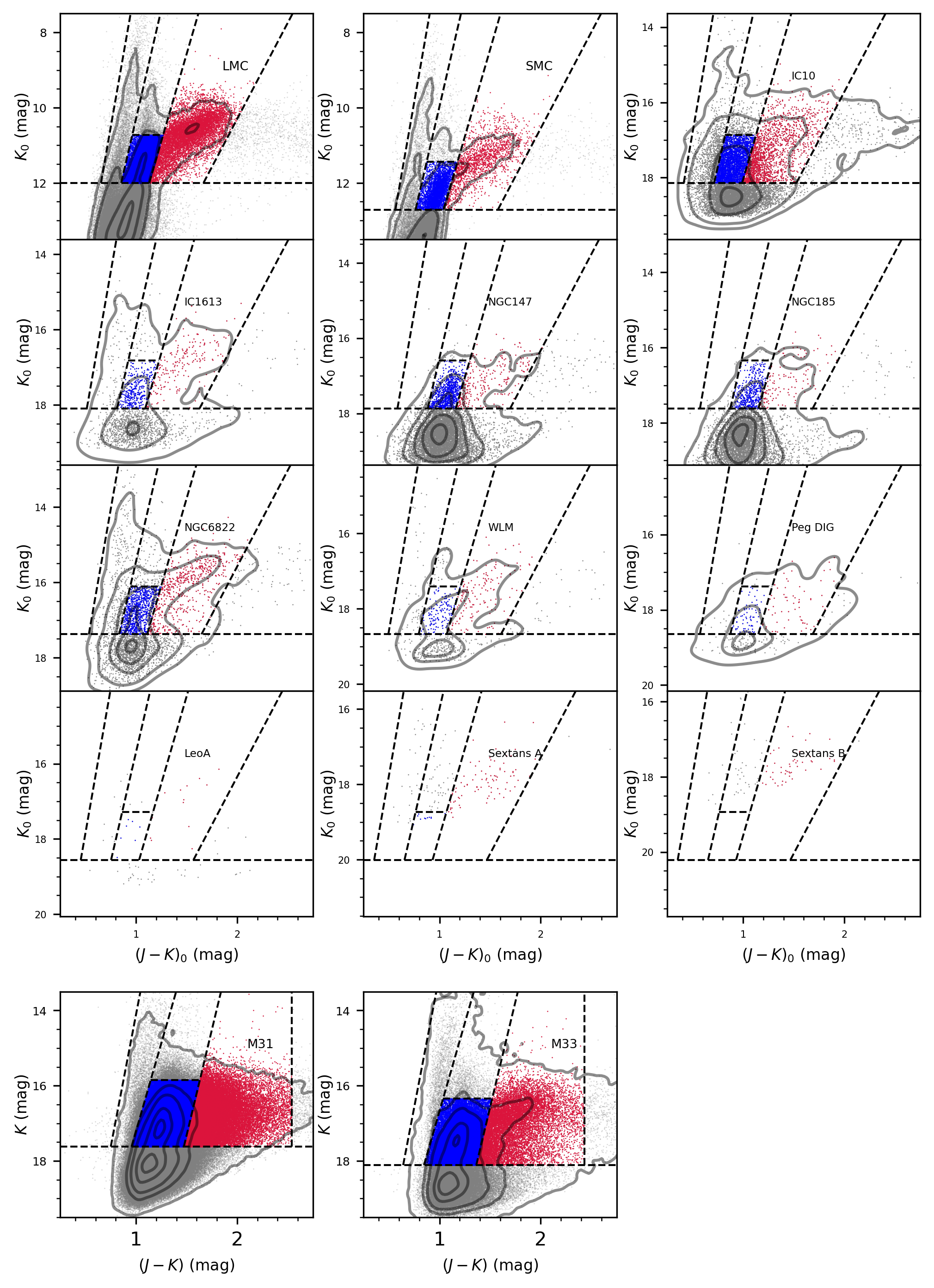}
    \end{adjustwidth}
    \caption{The classification by shifting the boundary lines of the LMC and M33, the benchmark galaxies, to each galaxy according to their positions of TRGB. The C-AGBs and O-AGs are represented by red and blue dots, respectively. Note that 12 galaxies, except M31 and M33, have been extinction corrected.
    \label{fig2}}
\end{figure}

With the borderlines of the LMC obtained above, the type of stars is identified in the sample of the SMC and the other ten dwarf galaxies, whose extinctions are corrected \mbox{in \citetalias{ren2021rsglocal}}, by shifting the borderlines on the CMD according to the position of their TRGBs relative to the LMC. Taking the SMC as an example, its borderlines are shifted from that of the LMC, by the difference in the TRGB location, i.e., $(J-K)_{0}^\mathrm{TRGB}=1.00$, $K_{0}^\mathrm{TRGB} = 12.00$ for the LMC to $(J-K)_{0}^\mathrm{TRGB}=0.91$, $K_{0}^\mathrm{TRGB} = 12.71$ for the SMC \citepalias{ren2021rsglocal}. The upper middle panel in Figure \ref{fig2} shows the borderlines for the SMC, which is in good agreement with its CMD structure.

In addition, a small modification is made for the ten dwarf galaxies. The blue borderlines of the RSG stars, i.e. the leftmost borderlines, are shifted bluewards by 0.1 mag to account for the greater dispersion caused by the larger photometric error at the faint ends than the LMC, which results in better agreement with the locations of the RSG in the $(J-K)_0/K_0$ diagram. The results are shown in Figure \ref{fig2}.

\subsection{The Borderlines of M31 and M33 \label{section34}
}
We study the samples of M31 and M33 separately. Their shapes on the CMDs look different from the above twelve galaxies whose borderlines are based on the LMC. For M33, its O-AGBs locate at a much wider range of $(J-K)$ color indices, which is particularly evident at the faint positions. A similar problem exists for M31, which neither shows a distinct C-AGB tail. Applying the LMC's borderlines directly cannot fit their actual CMD contours.

Considering this difference, we create a separate group of borderlines for M33 by the same method as for LMC, i.e., using {the GMM method}. As mentioned earlier, no extinction correction is performed for M31 and M33. Reference \citep{wang2021red} found that the extinction mostly ranges from $A_{\rm V} \sim 0.1$ to $0.15$ mag for M33, i.e., $A_{\rm K}\sim 0.01$ to $0.015$, very small compared to the photometric uncertainty. Though precise extinction correction would improve the distinction between different types of stars, the uncertainty in the extinction on the contrary can bring additional confusion and blur the borderline. \citet{wang2021red} applied the SFD98 extinction map to M31 and M33 and found that the extinction was generally over-estimated, which resulted in some fake features in the CMD in particular for M31 {($A_{\rm V} \sim 0.3$ to $1.3$, according to \citet{wang2021red}).}

Specifically, we select seven magnitude cuts in M33: four of them are concentrated near $K = 16.0$ mag: $K = 15.5$, 16.0, 16.4, and 16.6 mag. The three-component GMM is used to fit the number of sources at these four cuts, providing a reliable reference for separating C-AGBs and O-AGBs, with ppf = 0.01 and 0.99 limits for the blue-most RSGs and red-most C-AGBs, respectively. The single Gaussian distribution is taken at $K = 17.8$ and 17.0 mag, where ppf = 0.01 and 0.98 are set at the blue-most and red-most positions to constrain the large dispersion of the C-AGBs. Besides, ppf = 0.1 and 0.9 are set at $K = 14.5$ mag to constrain the luminous RSGs. Then the positions of Line 1, Line 2, Line 3 on the CMD of M33 are determined by linear fitting. Line 4 is taken from the average of several red limits, reaching $J-K = 2.431$. Line 5 is again detected by the Sobel filter as in the LMC with a brightness of $K = 16.338$. The lower panel of Figure \ref{fig1} shows the classification of M33, it can be seen that Line 3 has a larger uncertainty than the LMC in the faint area.

Afterwards, the borderlines for M31 are determined by shifting the TRGB of M31 relative to M33. Since the classification of stars in M33 is based on the sample without extinction correction, their TRGB positions are different from that in Table \ref{table1}. We take $(J-K)_\mathrm{TRGB}=1.20$, $K_\mathrm{TRGB} = 17.62$ for M31 and $(J-K)_\mathrm{TRGB}=1.09$, $K_\mathrm{TRGB} = 18.11$ for M33 according to \citetalias{ren2021red}, respectively. Like for the 10 dwarf galaxies, the blue borderlines of RSGs in M31 and M33, are shifted bluewards by 0.1 mag. The results are also shown in Figure \ref{fig2}.

The numerical slope and intercept for all the borderlines of the LMC and M33 are presented in Table \ref{table2}.

\begin{table}[H]
\caption{The coefficients of borderlines in the $(J-K)_0/K_0$ diagram of LMC and M33.  \label{table2}}
	\newcolumntype{C}{>{\centering\arraybackslash}X}
		\begin{tabularx}{\textwidth}{CCCCCC}
			\toprule
			{} & \textbf{Line 1} &  \textbf{Line 2} & \textbf{Line 3} &  \textbf{Line 4} &  \textbf{Line 5}\\
			\midrule
			LMC & $k=-15.366$ & $k=-11.618$ & $k=-9.268$  & $k=-5.111$ & {} \\
			{} & $b=22.048$ & $b=21.917$ & $b=22.481$ & $b=20.522$ & $y = 10.727$ \\ \midrule
			M33 & $k=-14.065$ & $k=-9.354$ & $k=-11.228$  & $x = 2.431 $ & {} \\
			{} & $b=28.526$ & $b=26.026$ & $b=33.416$ & {} & $y = 16.338$ \\
			\bottomrule
		\end{tabularx}
\end{table}

\section{Result and Discussion}  \label{section4}

\subsection{The Number of O-AGB and C-AGB Stars} \label{section41}

With the borderlines defined above, all the member stars are assigned to a certain class. The number of stars of each type is presented in Table \ref{table3}, including C-AGBs, O-AGBs, TP-AGBs, x-AGBs as well as RSGs. As noted by \citetalias{ren2021rsglocal}, the samples of Sextans A and B are incomplete because the observational sensitivity did not reach its TRGB brightness (c.f. Figure 7 of \citetalias{ren2021red}). Besides, the completeness of Leo A is slightly better while not enough to include all the O-AGBs. In the case of incompleteness due to insufficient sensitivity, the derived C/M ratio would be over-estimated since C-AGBs are brighter and thus more complete than O-AGBs. This bias is remarkably visible in the case of Sextans B where no O-AGB star is found at all (c.f. Table \ref{table3}). Although the RSG sample may be incomplete for the metal-poor galaxies due to wrong rejection  because their NIR colors can be as blue as foreground Galactic dwarf stars \citetalias{ren2021rsglocal}, this should not be a serious problem for AGB stars that is apparently redder. Consequently, the sample of AGB stars is basically complete for all the galaxies except Sextans A, Sextans B and possibly Leo A.
\begin{table}[H]
\caption{The C/M ratio, and number of various types of stars in the 14 galaxies\label{table3}}
	\begin{adjustwidth}{-\extralength}{0cm}
		\newcolumntype{C}{>{\centering\arraybackslash}X}
		\begin{tabularx}{\fulllength}{CCCCCCCCC}
			\toprule
		\textbf{Host galaxy }& \textbf{C/M ratio} & \textbf{C-AGBs} & \textbf{O-AGBs} & \textbf{TP-AGBs} & \textbf{x-AGBs} & \textbf{Others} & \textbf{RSGs} & \textbf{RSGs (\citetalias{ren2021red} \& \citetalias{ren2021rsglocal})} \\
			\midrule
			LMC       & $0.402^{+0.206}_{-0.054}$ & 8,160  & 20,283 & 2,041 & 1,359 & 193 & 4,578 & 4,823\\
			SMC       & $0.521^{+0.391}_{-0.101}$ & 1,769  & 3,396 & 526 & 302 & 96 & 2,032 & 2,138\\
			NGC 6822   & $0.674^{+0.301}_{-0.145}$ & 587 & 871 & 146 & 109 & 2 & 406 & 465\\
			NGC 185    & $0.289^{+0.438}_{-0.094}$ & 132 & 456 & 12 & 46 & 0 & 24 & 36\\
			IC 10      & $0.597^{+0.418}_{-0.169}$ & 1,840 & 3,083 & 309 & 286 & 9 & 1,103 & 1,340\\
			NGC 147    & $0.290^{+0.382}_{-0.088}$ & 264 & 910 & 20 & 106 & 3 & 48 & 82\\
			IC 1613    & $0.584^{+0.412}_{-0.129}$ & 198 & 339 & 39 & 35 & 1 & 100 & 115\\
			M31       & $0.165^{+0.227}_{-0.033}$ & 37,003 & 223,805 & 4,183 & 997 & 84 & 6,619 & 5,498\\
			Leo A     & $1.857^{+0}_{-0.635}$ & 13 & 7 & 2 & 7 & 0 & 10 & 10\\
			M33       & $0.341^{+0.306}_{-0.046}$ & 11,957 & 35,091 & 1,628 & 1,025 & 8 & 2,222 & 3,055\\
			WLM       & $1.158^{+0.647}_{-0.331}$ & 161 & 139 & 31 & 48 & 1 & 54 & 63\\
			Pegasus Dwarf & $0.861^{+0.316}_{-0.155}$ & 68 & 79 & 4 & 14 & 0 & 18 & 31\\
			Sextans A & $7.364^{+1.064}_{-1.364}$ & 84 & 11 & 43 & 5 & 0 & 30 & 33\\
			Sextans B & None & 49 & 0 & 34 & 2 & 0 & 12 & 14\\
			\bottomrule
		\end{tabularx}
	\end{adjustwidth}
\end{table}

\subsection{The Relation Between C/M Ratio and $(J-K)_\mathrm{TRGB}$} \label{section42}

The C/M ratio is the direct ratio of the number of C-AGB and O-AGB stars identified. Their relation with the color index $(J-K)_0$ of TRGB (derived in \citetalias{ren2021rsglocal}) is shown for all the sample galaxies but Sextans A and Sextans B in Figure \ref{fig2}. We present the C/M ratios as well as their uncertainties within the 50\% confidence interval for the fourteen galaxies defined in Section \ref{section3}. As shown in Figure \ref{fig1}, this asymmetric uncertainty may affect the value of the C/M ratio, and the C/M upper limit of individual galaxies, such as M31, can double the median value, which neither fit its actual shape on the CMD. So we ignore this asymmetry and take a linear fit that yields:
\begin{equation}
    \lg\mathrm{(C/M)} = -2.76\times (J-K)_0^\mathrm{TRGB}+2.44
\end{equation}

\begin{figure}[H]
    \includegraphics[width=13cm]{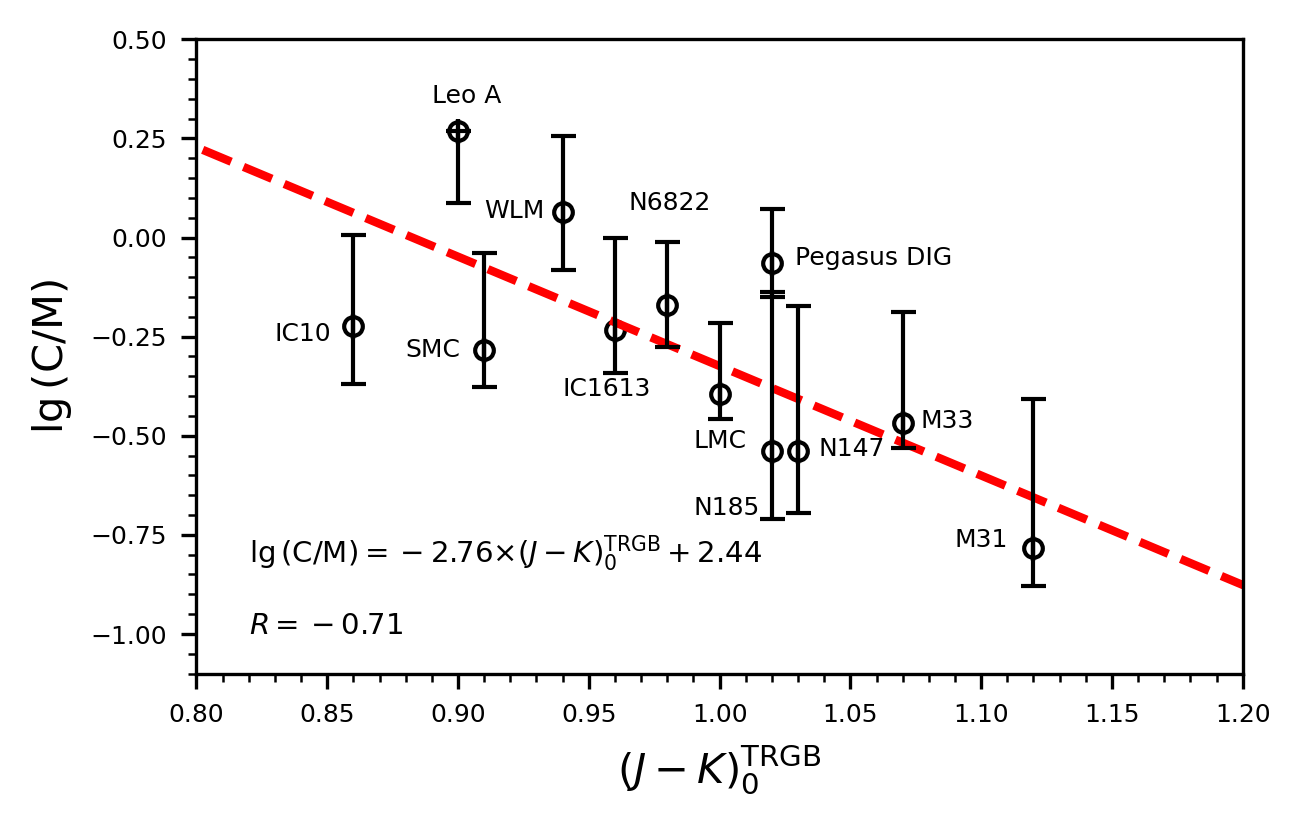}
    \caption{Dependence of $\lg(\mathrm{C/M})$ with $(J-K)_0^\mathrm{TRGB}$ for the 12 galaxies. The red dashed line represents the linear fitting whose result with the Pearson's correlation coefficient is displayed in the lower-left corner.
    \label{fig3}}
\end{figure}

\noindent{with Pearson's correlation coefficient of $R = -0.71$, indicating a certain correlation (Figure \ref{fig3}). It should be noted that the number of C-AGBs and O-AGBs and the ratio C/M would be little changed if the photometric accuracy is relaxed for LMC and SMC. Specifically, the numbers with no constraint on photometric quality lead to C/M=0.40 with 8,192 C-AGBs and 20,407 O-AGBs for the LMC, and C/M=0.52 with 1,788 C-AGBs and 3,431 O-AGBs for the SMC. This proves the completeness of the sample we select from the MCs. For the other distant galaxies, the sample of O-AGBs may not be so complete as the C-AGBs because C-AGBs are on average more than 1 mag brighter than the faint and numerous O-AGBs. In~such cases, the C/M ratio can serve as the lower limit. Nevertheless, \citetalias{ren2021rsglocal} demonstrated that the AGB stars are mostly complete in these galaxies.}

Moreover, the $J-K$ color index of TRGB is an indicator of metallicity. \citet{bellazzini2004calibration} established the relation between the absolute magnitude of TRGB in the $J$ and $K$ band with metallicity, from which the relation of $(J-K)_0$ with metallicity can be derived as: 
\begin{equation}
     (J-K)_0^\mathrm{TRGB}=0.38\mathrm{[M/H]}+1.38
\end{equation}
which generally agrees with that derived by \citet{ivanov2000extending} for RGB stars.

With this relationship, the above equation on the C/M ratio can be converted to:
\begin{equation}
    \lg\mathrm{(C/M)} = -1.04\times\mathrm{[M/H]} - 1.37.
\end{equation}
Alternatively, the metallicity can be derived from the C/M ratio:
\begin{equation}
    \mathrm{[M/H]}= -0.96 \lg\mathrm{(C/M)} - 1.32
\end{equation}

This can be compared with the relation derived by \citet{battinelli2005calibration} and refined by \citet{cioni2009metallicity}:
$ \mathrm{[Fe/H]} = -0.47 \lg(\mathrm{C/M}) - 1.39 $, showing good consistency. It should be noted that the metallicity derived in this way might be lower for metal-rich galaxies. For example, the metallicity would be -0.87 for M33, much lower than the spectroscopically derived value. This originates from  the relation between the metallicity and color index $J-K$, which needs further refinement.

\subsection{Comparison with Previous Works} \label{section43}

The previous measurements of C/M ratio are inhomogeneous, and different even for the same galaxy. Some typical results  in recent years are listed in Table \ref{table4} where one galaxy is selected with only one representative result (e.g., the C/M ratio of NGC 6822 has been measured for a few times with various values, 0.23 in \citet{cioni2005near}, 0.27 in \citet{kang2006evolved}, 0.62 in \citet{sibbons2012agb}, 0.53 in \citet{hirschauer2020dusty}, and the latest one is selected in Table \ref{table4}). For M31, the result was based on the observation of part of the galaxy, i.e five square regions by the narrow-band photometry ([TiO]-[CN]), which can be very different from the present study in completeness and preciseness. For the other galaxies, the CMD method is used to identify O-AGBs and C-AGBs, which follows the same principle of this work in spite of different bands and/or different depths. The different results come from a few factors, mainly the pureness and completeness of the sample, as well as the borderlines. This will lead to a lack of homogeneity in previous C/M ratio studies. While our work improves this deficiency based on homogeneous data and semi-automatic classification methods. We remove foreground dwarfs with the $J-H/H-K$ diagram and the astrometric measurement from GAIA in our sample. Besides, the borderlines of the galaxies in Section \ref{section3} are shifted based on the benchmark galaxies considering both the variation of the brightness and $J-K$ color index of TRGB, as the $(J-K)_\mathrm{TRGB}$ characterizes the metallicity differences of the AGB stars. In addition, the photometric error is limited to less than 0.05 mag for Magellanic Clouds and 0.2 mag for distant galaxies to keep the consistency of our sample.

\begin{table}[H]
\caption{The C/M ratio in previous studies. \label{table4}}
	\begin{adjustwidth}{-\extralength}{0cm}
		\newcolumntype{C}{>{\centering\arraybackslash}X}
		\begin{tabularx}{\fulllength}{CCCCCCC}
			\toprule
			\textbf{Host Galaxy} & \textbf{C/M Ratio} & \textbf{C-AGBs} & \textbf{O-AGBs} & \textbf{Data Resource} & \textbf{Method} & \textbf{References}\\
			\midrule
			LMC       & 0.3 & 7,572  & 25,229 & DENIS & $(J-K_s)/K_s$ CMD & \citet{cioni2003agb}  \\
			SMC       & 0.27 & 1,643 & 6,009 & DENIS & $(J-K_s)/K_s$ CMD & \citet{cioni2003agb} \\
			NGC 6822   & 0.53 & 560 & 1,050 & UKIRT & $(J-K)/K$ CMD & \citet{hirschauer2020dusty} \\
			NGC 185    & 0.11 & 73 & $\sim$660 \textsuperscript{1} & CFHTIR & $JHK'$ CMD and CCD & \citet{kang2005near} \\
			IC10      & 0.30 & 531 & 1,766  & HST WFC3/IR & (F127M-F139M)/(F139M-F153M) CCD & \citet{boyer2017infrared} \\
			NGC 147    & 0.25 & 65 & 265 & HST WFC3/IR & (F127M-F139M)/(F139M-F153M) CCD & \citet{boyer2017infrared}\\
			IC 1613    & 0.52 & 291 & 552 & UKIRT & $(J-K)/K$ CMD & \citet{sibbons2015agb} \\
			M31       & 0.084 & 945 & 11,228  & CFH12K & [TiO]-[CN] method & \citet{battinelli2003carbon}  \\
			Leo A     & 1.80 & 18 & 10 & WHIRC & $J$ - [3.6]/[3.6] CMD & \citet{jones2018near} \\
			M33       & 0.35 & 7,404 & $\sim$21,150 \textsuperscript{1} & UKIRT & $(J-K_s)/K_s$ CMD & \citet{cioni2008agb}  \\
			WLM       & 0.56 & 146 & 259 & NTT SofI & $(J-K_s)/K_s$ CMD & \citet{valcheva2007carbon} \\
			Pegasus Dwarf & 1.38 & 44 & 32 & HST WFC3/IR & (F127M-F139M)/(F139M-F153M) CCD & \citet{boyer2017infrared} \\
			\bottomrule
		\end{tabularx}
	\end{adjustwidth}
	\noindent{\footnotesize{\textsuperscript{1} The number of O-rich AGB stars of M33 and NGC 185 is estimated from the number of C-AGBs and the C/M ratio.}}
\end{table}

\begin{figure}[H]
\begin{adjustwidth}{-\extralength}{0cm}
    \centering
    \includegraphics[width=19cm]{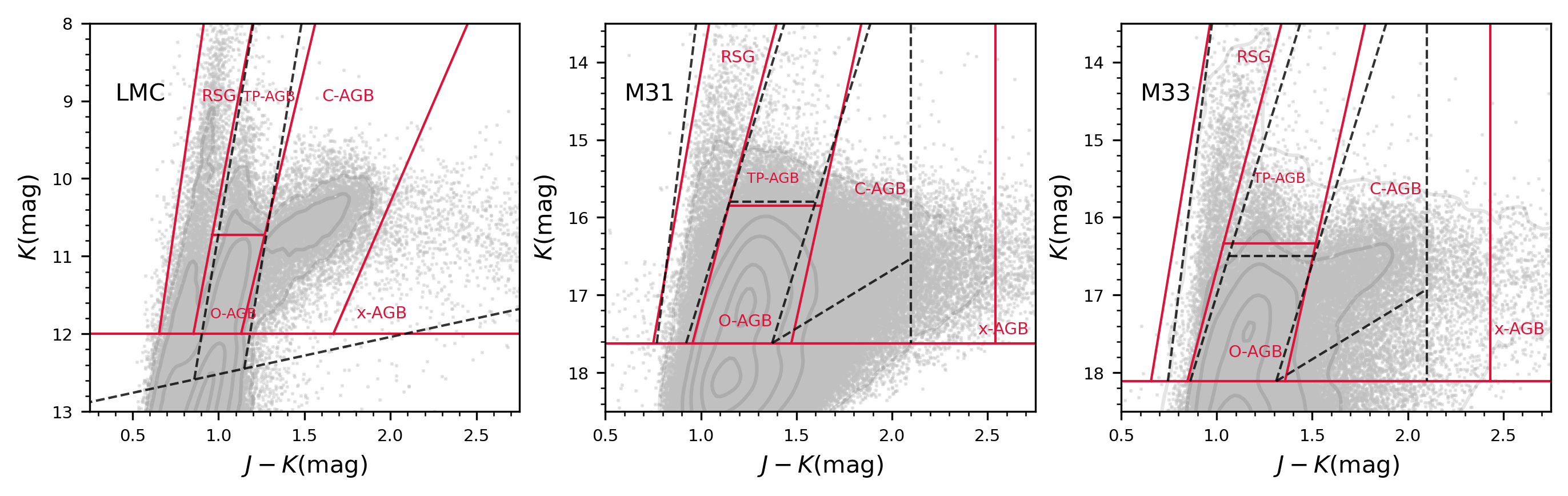}
    \end{adjustwidth}
    \caption{Comparison of the borderlines with previous studies, where the red solid lines represent this work and the grey dashed lines represent the work by \citet{cioni2006agb} for the LMC and by \citetalias{ren2021red} for M31 and M33.
    \label{fig4}}
\end{figure}

\textls[-15]{\citet{cioni2003agb} selected the AGB stars from the $I/I-J$ CMD for the Magellanic Clouds, because the TRGB position in $I$ band  is generally a constant value \citep{freedman2020calibration}. Then, the separation of C-AGB and O-AGB stars was set at $J-K_s=1.4$.\mbox{ Similarly, \citet{cioni2005near}}} took $J-K_s=1.36$ for NGC 6822, and \citet{sibbons2012agb} took $J-K_s=1.28$ for the same galaxy to divide the C- and O-AGB stars. This difference comes from a few factors, e.g., the contamination from foreground stars, different photometric accuracy and subjective selection. Apparently, a vertical line in the CMD is not robust enough to separate the C-AGB tail on the CMD. In Figure \ref{fig4} (left panel), the borderlines of \citet{cioni2006agb} (grey dashed line) are plotted for comparison, for which the interstellar extinction is corrected by shifting 0.049 mag leftward and 0.023 mag upward according to the color excess E(B-V) in \citetalias{ren2021rsglocal} and the extinction law of \citet{wang2019optical}. The slightly moved borderlines agree reasonably well with the slopes of O-AGBs in this work. An evident difference is that classification from \citet{cioni2006agb} has a lower limit for O-AGBs fainter than the TRGB, which increases the number of O-AGBs and can account for their lower {C/M} ratio for the MCs. Besides, we divide AGB stars into more sub-types including TP-AGBs and x-AGBs that are not considered in \citet{cioni2006agb}.

This work uses the sample of \citetalias{ren2021red} for M31 and M33, but the numbers of C-AGBs and O-AGBs and consequently the C/M ratio are different. However, this difference is not because of the contamination or incompleteness as mentioned above. Instead, the comparison shown in Figure \ref{fig4} (the middle and right panels) clearly illustrates that it is caused by the re-defined borderlines. It can be seen that the classifications of RSGs, O-AGBs, and TP-AGBs of M31 and M33  are very similar in our results and \citetalias{ren2021red}, and the biggest difference comes from the C-AGBs. Due to the considerable  extension of the C-AGB tail of these two spiral galaxies, we set a red limit line at $(J-K)\sim2.5$ as the borderline between C-AGBs and x-AGBs to increase the completeness of the C-AGBs. In addition, the unclassified stars in the triangular area fainter than the C-AGBs in \citetalias{ren2021red} are also almost classified as C-AGBs. {As a result, the number of C-AGBs in M33 is increased from 10,218 to 11,957, compared to \citetalias{ren2021red}. However, for M31,
it is decreased from 46,692 to 37,003, because some stars labeled as C-AGBs in \citetalias{ren2021red} are now classified as O-AGBs  in this study. This is more consistent with the actual contour (the middle panel of Figure \ref{fig4}), and leads to the opposite trend.} It deserves to mention that the photometric uncertainty is relatively large due to their large distance and crowding. A high spatial resolution observation in space can be very helpful to confirm the identification of sources in M31 and M33.

The samples of twelve low-mass galaxies are from \citetalias{ren2021rsglocal} and the number of RSGs is in agreement with \citetalias{ren2021rsglocal} as shown in Table \ref{table3}. This fact confirms the reliability of the borderlines given in \citetalias{ren2021rsglocal} based on eye check from the LMC. Meanwhile, the current sample of RSGs is smaller, this is caused by removing the extended sources identified in the photometry procedure. Correspondingly, the number of any type of stars is reduced due to the shrinking of the whole sample. {The~difference for M31 and M33 appears to be greater. In particular, M31 has more RSGs than in \citetalias{ren2021red} (Table \ref{table3}), which can be explained by the redder borderline between dark RSGs and O-AGBs (Figure \ref{fig4}) that causes the sources previously classified as O-AGBs to be RSGs now. Nevertheless, due to the huge number of O-AGBs (223,805) in M31, this quantity change of $\sim$ 1000 has no significant impact on the C/M ratio.}

\subsection{The Uncertainties Analysis}
\label{section44}

We have selected the most typical sample of C-AGBs and O-AGBs on CMD to derive the C/M ratio. As the photometric error would move the stars into both sides of the borderlines and the stars are not symmetrically distributed, the sample can be contaminated by such an effect and the C/M ratio can be changed. In addition, the atypical AGB samples, such as TP-AGB tail and x-AGB stars, affect the ratio as well. The sloped TRGB may be another factor to change the ratio. Besides, the spectroscopically confirmed stars may actually locate in the adjacent regions of the CMD, blurring the borderline. We discuss these problems as follows.

\subsubsection{The Photometric Error}
\label{section443}

The photometric error can lead the C-AGB stars to the O-AGB area, and vice versa. Such effect of the photometric error was already considered to calculate the AGB pollution rate of RSG in Section 5.5.2 in \citetalias{ren2021rsglocal}. The basic idea is to build a model CMD with theoretical C-AGBs and O-AGBs with the typical observed photometric error at the corresponding magnitude, and then to count the number of AGBs that move to the area of RSGs. Here, the same method is used to calculate the effect of photometric errors on the C/O-AGB ratio by taking the benchmark galaxies LMC and M33 as examples. The~metallicity [M/H] is set to $-$1.0 for LMC and 0.0 for M33\endnote{Significantly higher than $-$0.82 in Table \ref{table1}, which is because the ${(J-K)}_{0}^\mathrm{TRGB}$ - [M/H] relation \mbox{in \citet{bellazzini2004calibration}} is calibrated based on the metal-poor environment of the globular clusters, and the [M/H] derived from this relation is too low when applied to the metal-rich environment.}, respectively.

The simulation of the LMC shows the high purity of the C- and O- AGBs. At $\mu$ (distance modulus) of 18 and 18.5, 93\% of the O-AGBs and 94\% of the C-AGBs are in the expected CMD areas, and they slightly decrease to 92\% and 93\% at  $\mu =19$, indicating that the vast majority of the AGBs are correctly classified. As for M33, the much further benchmark galaxy, the simulation shows that 97\% O-AGBs and 98\% C-AGBs can be correctly classified by our borderlines at $\mu =24$, while at $\mu=25$, the above numbers drop to 88\% and 96\%. These results demonstrate that the photometric error can lead to a higher C/M ratio but generally by less than 5\%.

\subsubsection{The TP-AGB tail}
\label{section441}

In the study of TP-AGB population in the MCs by \citet{pastorelli2019constraining,pastorelli2020constraining}, most of the O-AGBs are located in the middle area approximately parallel to both RSGs and C-AGBs on the CMDs. {However, there are also some unusually bright AGB stars that draw a ``tail'' adjacent to the RSGs. They are classified as O-AGBs in their study.} In \citetalias{ren2021rsglocal}, they are labeled as TP-AGB stars by the MIST model for stars of 5$M_{\odot}$ or 7$M_{\odot}$. Since TP-AGBs are in an evolutionary stage with rapidly increasing carbon content, the chemical type of these sources needs to be further analyzed. Here we introduce the Padova model \citep{bressan2012parsec,marigo2013evolution,tang2014new,chen2014improving,chen2015parsec,marigo2017new,pastorelli2019constraining} to analyze the properties of TP-AGB tail in the LMC as an example.

Here we demonstrate mainly the discrimination between O-AGBs and C-AGBs. The~Padova stellar isochrones of LMC are calculated with the absolute number of stars per unit mass based on the initial mass function \citep{kroupa2001variation,kroupa2002initial} and the composition of circumstellar dust of O-AGBs and C-AGBs being no dust and 100\% AMC \citep{groenewegen2006mid}, respectively. The~metallicity of the LMC is set to [M/H] $=-1.00$ (Table \ref{table1}). The distance modulus is set at 18.477 for the LMC \citep{pietrzynski2019distance}. The 1 Gyr isochrones {(blue dots)} have a good agreement with the CMD contour of LMC, but are fainter than the {TP-AGB tail}. While the relatively young and metal-rich 0.63 Gyr model shows that O-AGBs (green dots) match the TP-AGB tail. These younger AGB stars are more metal-rich and have lower 3DU efficiency. Therefore, they require larger mass and higher temperatures to form C-AGBs. This means that more metal-rich and young O-AGBs can move to the brighter area on the CMD than the more metal-poor and older AGB stars, and their evolutionary tracks (green dots) can explain the TP-AGB tail. Besides, the overall number of sources classified as TP-AGB stars is lower than 20\% of the O-AGB stars in the host galaxy (Table \ref{table3}), so the potential incompleteness of the O-AGB sample is limited.

\begin{figure}[H]
    \includegraphics[width=10cm]{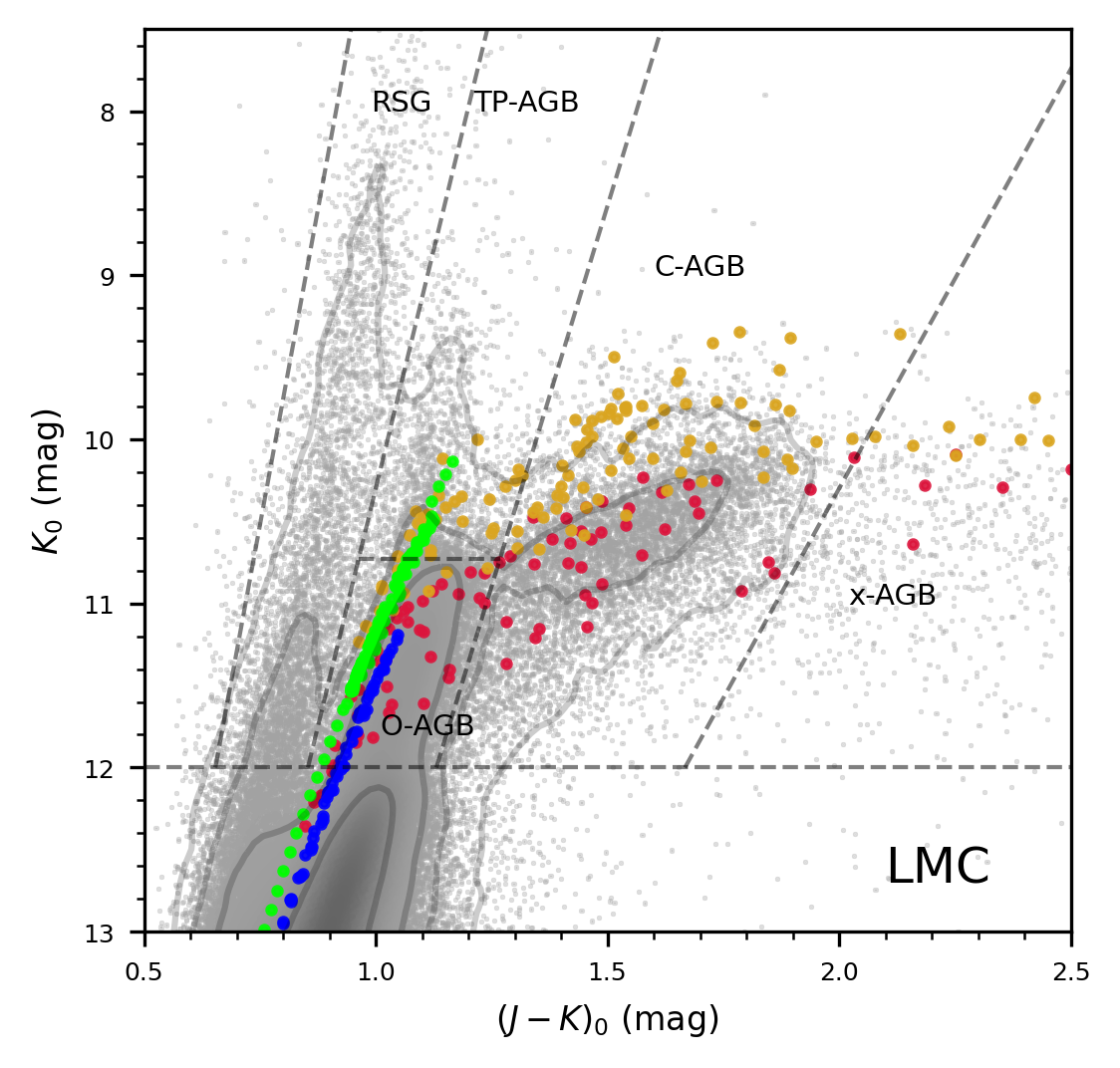}
    \caption{Verification of the classification areas in Figure \ref{fig1} with Padova stellar evolutionary models.  The red and blue dots represent C-AGBs and O-AGBs at an age of 1.00 Gyr, respectively, and the yellow and green dots represent C-AGBs and O-AGBs at an age of 0.63 Gyr, respectively. The  background is the sample stars in the LMC from \citetalias{ren2021rsglocal}.
    \label{fig5}}
\end{figure}

Figure \ref{fig5} also reflects the fact that some C-AGBs (red and yellow dots) are actually located in the area classified as the O-AGBs area in the CMD. This is related to the temperature increase of C-AGBs during the thermal pulse {process} \citep{marigo2017new}. The CMD region where the O-AGBs are located is subject to contamination by C-AGBs, implying the incompleteness of C-AGBs. In Section \ref{section443}, we illustrate this uncertainty by comparing it with the spectroscopically validated AGB star sample of LMC and NGC 6822.

\subsubsection{The Location of the C-AGB Stars}
\label{section442}

The spectroscopic classification is induced to estimate the level of the C-AGBs mixed into the O-AGBs area. The AGB sample in three 0.12 square degrees areas surveyed by \citet{blanco1980carbon} with the CTIO 4m telescope is used in our evaluation. There are 186 C-AGBs in this spectroscopic sample, 90\% of them locate in the C-AGBs area of the CMD, but 14 locate in the O-AGBs area, and a few are below the TRGB. As for 102 O-AGBs, almost all of them are located in the corresponding area of the CMD. The spectroscopic types of the AGB stars in the TP-AGB area are identified as almost all O-rich, which is consistent with the discussion in Section \ref{section441} that TP-AGB stars in the LMC are mainly oxygen-rich.

\begin{figure}[H]
    \includegraphics[width=14cm]{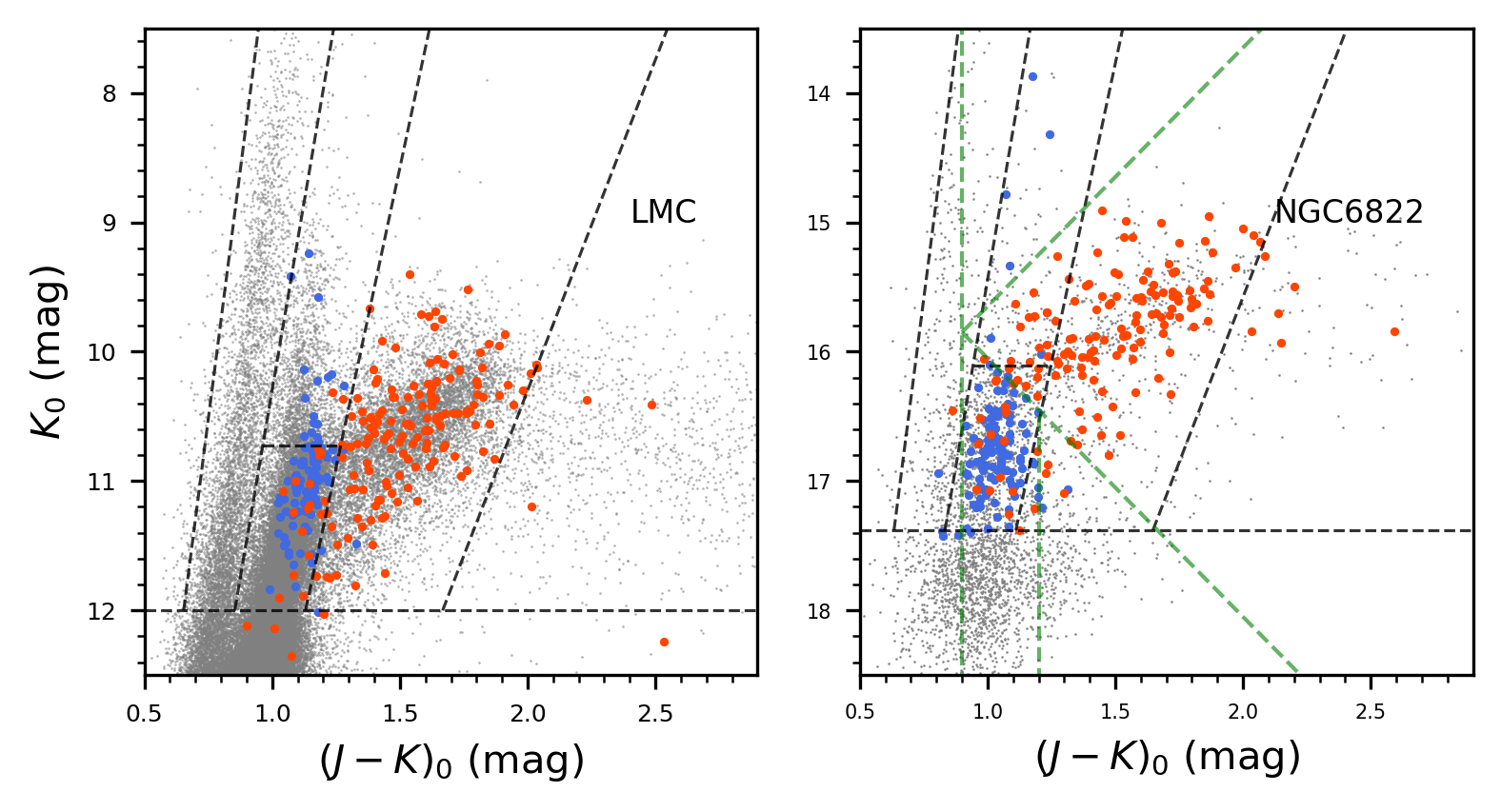}
    \caption{The previous spectroscopic classifications of AGB stars in the LMC (left) and NGC6822 (right) are used for the validation of our $J/J-K$ CMD photometric method. The C-AGBs stars and O-AGBs identified from previous spectral observations are indicated by red and blue dots, respectively. The black dashed line represents the borderlines in our work, and the green dashed line in the right panel represents the borderlines suggested by \citet{kacharov2012spectra} based on the result of the NGC 6822 AGB stars' spectroscopic analysis.
    \label{fig6}}
\end{figure}

The metal-poor galaxy NGC 6822 is also verified by the spectroscopic result. We~chose the NGC 6822 AGB spectroscopic classification of \citet{kacharov2012spectra} with the VLT's VIMOS instrument. This sample contains 546 AGB stars, including 151 C-AGBs more sparsely distributed in the CMD area than the C-AGBs in the LMC, and \citet{kacharov2012spectra} defined a group of borderlines accordingly (right panel in Figure \ref{fig6}). Compared to their classification, our borderlines can ensure that the vast majority of the spectroscopically verified O-AGBs are located at the photometric classified O-AGB area of the CMD, and 107 C-AGBs are in the expected position on the CMD. However, the other one-third of C-AGBs deviate from the expected location, 20 in the O-AGBs area, and 16 in the TP-AGB area. Curiously, NGC 6822 is relatively metal-poor with better 3DU efficiency so its C-AGBs should be fainter \citep{marigo2017new} and the TP-AGB area of NGC 6822 should be dominated by O-AGBs, but in fact, most of the stars there are C-AGBs. This anomalous distribution deserves further study.

In conclusion, the examples of the LMC and NGC 6822 show that O-AGBs are indeed concentrated in the corresponding area on the CMD, while C-AGBs may locate in other areas on the CMD and cause contamination, resulting in an underestimation of the number of C-AGBs in our work. The existing spectroscopic classifications in other galaxies suffered from a lack of data, inhomogeneity and incompleteness, and these uncertainties vary across galaxies. Meanwhile, the stars that locate in the TP-AGB area of the CMD are not certainly dominated by O-AGBs.

\subsubsection{x-AGB Stars}
\label{section444}
\textls[-25]{Another factor that may lead to underestimating the number of C-AGB stars is the presence of x-AGB stars. They are dusty, unusually red, and most of the carbon-rich type \citep{blum2006spitzer,boyer2011surveying,dell2015agb}}. In~Section \ref{section3}, we set the red limits of $J-K$ for C-AGBs, resulting in the number of x-AGBs being below 20\% of the overall number of C-AGBs. x-AGBs may introduce uncertainties in the final results of the C/M ratio. However, the incompleteness brought by the x-AGBs to C-AGBs and the incompleteness brought by the TP-AGB tail to O-AGB stars to the C/M ratio cancel each other to some degree. We have selected the vast majority of the typical C-AGBs and O-AGBs on the CMD. Moreover, as shown in the results of Section \ref{section42}, the C/M ratio vs $(J-K)_0^\mathrm{TRGB}$ relation is well defined.

\subsubsection{The Tip of RGB}
\label{section445}

\begin{figure}[H]
    \includegraphics[width=10cm]{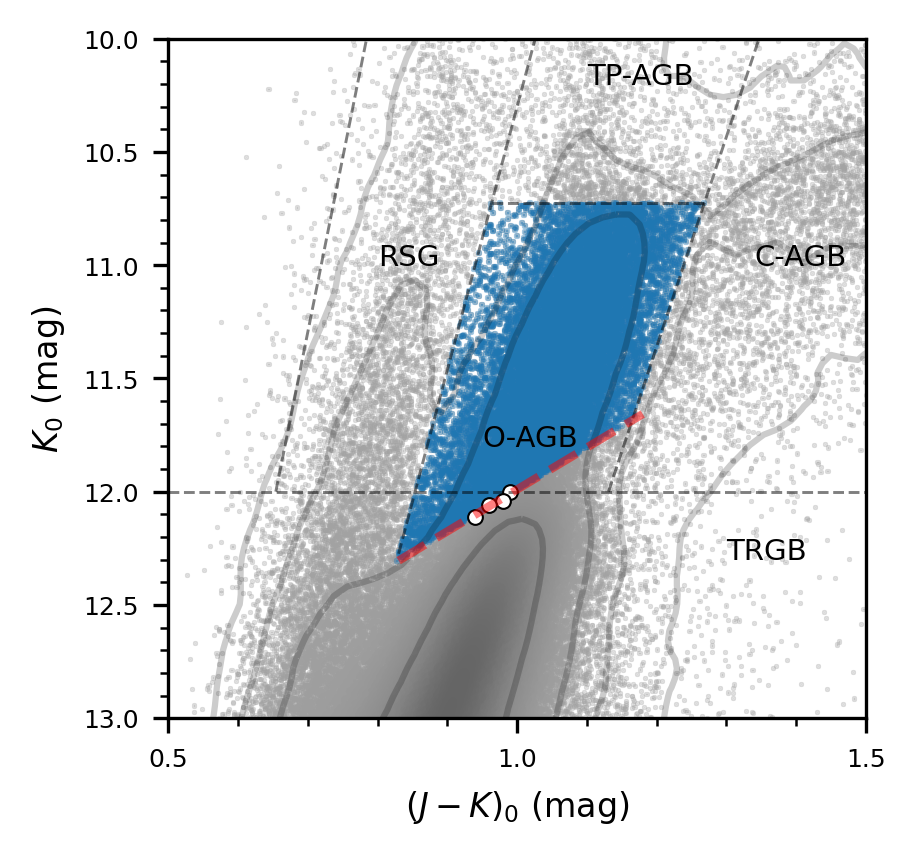}
    \caption{Comparison of the results of the LMC taking sloped TRGB and horizontal TRGB. The~red line represents the sloped TRGB, and the white dots represent the saddle points we detected in different bins by the ridge method of number density analysis. The O-AGBs suggested by the sloped TRGB are represented by blue dots.
    \label{fig7}}
\end{figure}

The TRGB of $K$ band is not actually horizontal on the CMD, but has an age-dependent slope \citep{serenelli2017brightness,hoyt2018near,freedman2020calibration}. Nevertheless, the Padova stellar model suggests that more than 90\% of the TP-AGB stars is above the TRGB \citep{marigo2008evolution,boyer2015identification}. Actually, in many previous studies of the C/M ratio, the classification of AGB stars only considered stars brighter than the horizontal TRGB (e.g., \citet{sibbons2012agb,hirschauer2020dusty}). In \citetalias{ren2021red} and \mbox{\citetalias{ren2021rsglocal}}, we determined the TRGB brightness of the LMC to be $K = 12.00$ by searching for the saddle point on the 2D density distribution on the CMD. In order to determine how much the slope of the TRGB has an impact, we selected the density saddle points at the neighboring bins, i.e., at $(J-K)_0$ of 0.94, 0.96, 0.98 and 0.99. Afterwards, a linear fit of these saddle points yields a sloped TRGB representing the number density mutation of bright RGB and faint O-AGBs on the CMD (Figure \ref{fig7}):

\begin{equation}
    K = -1.85\times(J-K)_0 + 13.84
\end{equation}

When the TRGB of the LMC is changed from $K = 12.00$ to this sloped one, the number of O-AGBs slightly changed from 20,283 to 20,289. This indicates that the adoption of a horizontal TRGB has little effect on the C/M ratio and it is consistent with the experience of previous studies.
\subsection{The Radial Distribution of the C-AGB, O-AGB Stars and C/M Ratio in M31 and M33} \label{section45}

\begin{figure}[H]
  \begin{adjustwidth}{-\extralength}{0cm}
      \centering
      \includegraphics[width=15cm]{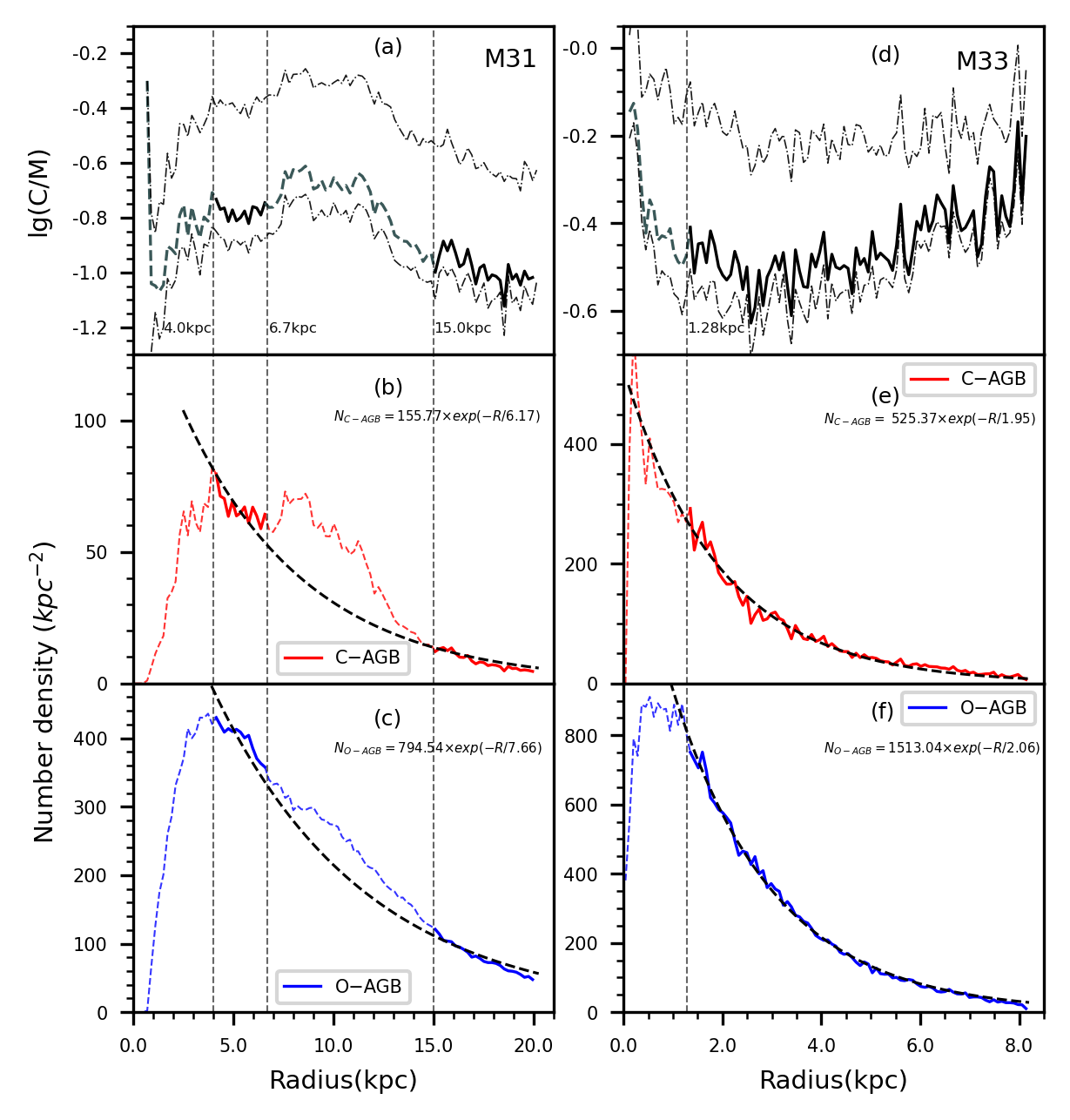}
      \end{adjustwidth}
      \caption{The radial distribution of the C/M ratio, C-AGB stars and O-AGB stars in M31 (panel a, b, c), and in M33 (panel d, e, f). An exponential function is used to fit the radial distribution of C-AGBs and O-AGBs with the results displayed, where the region with uncertain photometry and locate in the ring area in M31 are not used and labelled by the dashed line. {Besides, the 50\% confidence limits of the C/M ratio are represented with dashed-dot lines on both sides of the C/M ratio median lines.}
      \label{fig8}}
  \end{figure}

For the spiral galaxies M31 and M33, the radial distribution of the C/M ratio is investigated. The area of study is taken from \citetalias{ren2021red} defined by \mbox{B = 25 mag/arcsec$^2$} isophote, which corresponds to an ellipse with the major and minor axis of \mbox{20.26 kpc $\times$ 6.56 kpc} for M31 and 8.18 kpc $\times$ 4.82 kpc for M31. Both ellipses are divided equally into 100 equal-width bins along their axes. The radial distribution along the major axis is displayed in Figure \ref{fig8} for the C-AGBs, O-AGBs and C/M ratio, respectively.

The distribution of surface density in number/kpc$^2$ is fitted by an exponential function. Because the photometry in the very inner part can be poor due to crowding and O-AGBs can be more affected due to their fainter brightness than C-AGBs, the data with the major axis radius $r<4.0$ kpc for M31 and $r<1.28$ kpc for M33 are discarded in the fitting. In~addition, the evident ring of M31 from $\sim 6.7$ kpc to 15.0 kpc is discarded in our fitting as well because of the apparently higher stellar density. The fitting yields a radial scale length of 6.17 kpc and 7.66 kpc for C-AGB and O-AGB stars, respectively, in M31, while 1.95 kpc and 2.06 kpc, respectively ,in M33. Both show that the radial scale length of O-AGBs is larger than of C-AGBs, which means O-AGBs distribute into a more extended space.

The C/M ratio exhibits a contrary tendency for M31 and M33 with the radius. {It is notable to demonstrate that the tendency is reliable, because even taking into account the uncertainty from the GMM model, the tendency at the 50\% confidence level still coincides with the median distribution. As can be seen in Figure \ref{fig3} and Table \ref{table3}, the upper limit deviates much more significantly because the borderline will locate at the high-density area of pre-assumed O-AGBs if the borderline moves blue-ward, and the number of C-AGBs increases considerably to bring about the larger C/M ratio. On the contrary, this effect is not so severe as the borderline moves red-ward to the low-density area.} In M31, the ratio decreases with the radius, while it increases with the radius in M33. For M33, this tendency coincides with the prediction of the inside-out model of galactic chemical evolution that describes the relationship of metallicity with radius in the disk of a spiral galaxy. Besides, \mbox{\citet{cioni2008agb} also} yielded an increasing C/M ratio with radius in M33. This is expected from the stellar evolutionary scenario that metal-poor stars can more easily evolve into C-AGBs than metal-rich ones.

However, the C/M gradient in M31 contradicts our expectation. In the previous study of \citet{hamren2015spectroscopic}, its C/M ratio was found to increase outward in the disk of M31 using a combination of moderate-resolution optical spectroscopy from the Spectroscopic Landscape of Andromeda's Stellar Halo survey (SLASH) and six-filter Hubble Space Telescope photometry from the Panchromatic Hubble Andromeda Treasury survey (PHAT). Considering that the carbon stars in this study were identified spectroscopically and consisted of only about 100 objects, the sample is far from complete. Besides, the contamination of foreground stars can induce further uncertainty. With a rather pure and complete sample of C-AGBs and O-AGBs as well as the removal of foreground stars in this work, the trend of C/M ratio decreasing with radial distance in M31 should be real. Moreover, this ratio exhibits an obvious increase within the M31's 11 kpc ring, i.e. between 7 and 15 kpc. One possibility is that the metallicity increases with galactocentric radius. However, \citet{pena2019metallicity} studied the metallicity in the disk of M31 by measuring planetary nebulae (PNe) and found that metallicity is almost independent of galactocentric radius. From the standpoint of stellar evolution, the stellar mass in addition to metallicity is another factor to determine whether a star can become C-AGB. 

\begin{figure}[H]
\begin{adjustwidth}{-\extralength}{0cm}
    \centering
    \includegraphics[width=10cm]{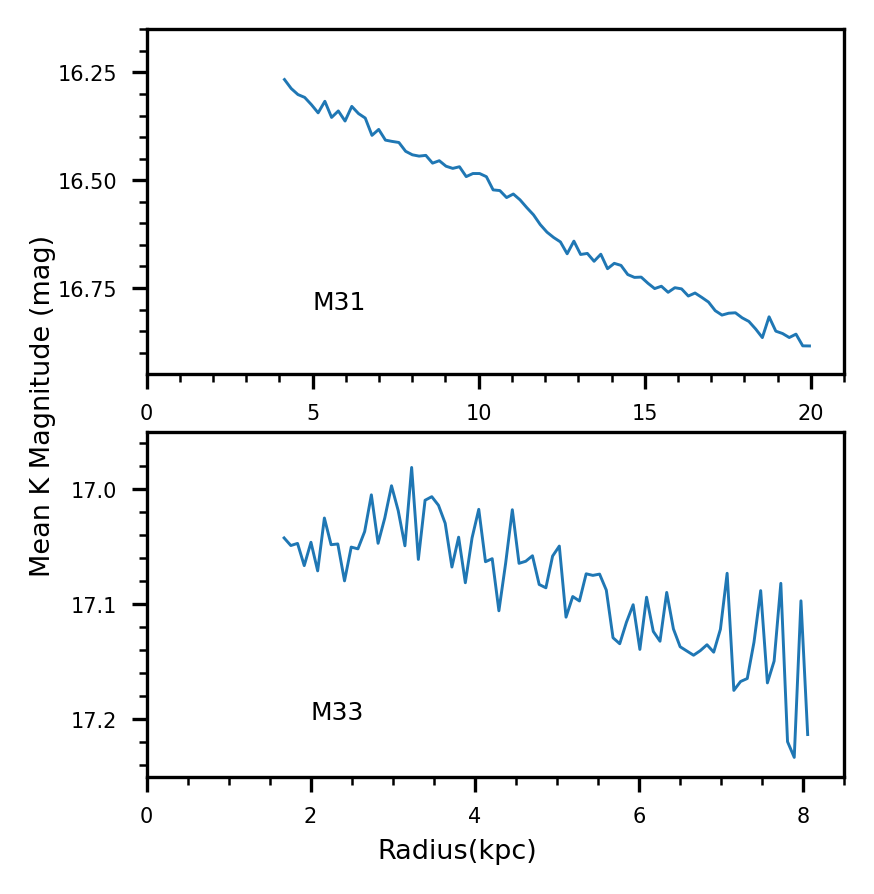}
    \end{adjustwidth}
    \caption{{The radial distribution of the mean K magnitude of the sources labelled as C-AGBs or O-AGBs in M31 (upper) and M33 (bottom).}
    \label{fig9}}
\end{figure}

As mentioned in Section \ref{section1}, a more massive star is easier to become carbon-rich. {We also analyze the $K$ band luminosity gradient of sources labeled as C-AGBs or O-AGBs in 100 equal-width bins. To ensure the reliability of the gradient, we adjust $\sigma_{JHK}$ from 0.2 mag to more strict 0.05 mag, and discard the innermost 20\% part when derive the mean $K$ magnitude of the AGB stars, to control the effect from the sources with relatively poor photometric accuracy, then the numbers of sources in M31 and M33 drop to 81,762 and 23,973, respectively. We find that both M31 and M33 have a decreasing gradient of brightness (Figure \ref{fig9}), and M31 is obviously more decreased at $\sim$ 0.5 mag at R = $\sim$ 5 kpc to $\sim$ 20 kpc, while this change in M33 is $\sim$ 0.2mag. This brightness trend implies that the stellar mass of the AGB stars in M31 decreases with increasing radius, resulting in an age gradient of the AGB stars \citep{feast2010there}. The lower mass of the AGB star reduces the 3DU efficiency, this fact can expalin the opposite C/M ratio tendency of M31. As shown in Figure \ref{fig9}, at \mbox{R = 5--15 kpc} where the luminosity gradient decreases not so much, there is an increase in C/M ratio in M31. The impacts from this gradient are limited for 3DU, and metallicity still dominates the evolution of AGB stars here, leading to the increasing C/M ratio there like M33. However, for where R $>$ 12 kpc, the brightness of the sources labelled as AGB stars further decreases. So the mass and age are replaced the metallicity factor, resulting in a decrease in the C/M ratio the outer regions of M31.}
However, the mass function depends on IMF and the star formation history, which deserves further study.

\section{Summary}  \label{section5}

The number ratio of C-rich to O-rich AGB stars has been studied previously in a few Local Group Galaxies, which generally suffer from incompleteness, contamination by foreground dwarf stars and inhomogeneity. Here we re-calculate the C/M ratio in 14 galaxies of the Local Group by the same method with the sample of AGB stars identified by \citetalias{ren2021red} and \citetalias{ren2021rsglocal}, for which the foreground stars are mostly removed with the near-infrared color-color diagram and color-magnitude diagram as well as the Gaia astrometric information. The samples have relatively higher completeness, pureness and homogeneity.

The borderlines that divide the C-AGB and O-AGB stars, as well as RSG stars, are determined for the benchmark galaxies LMC or M33 with the best-assembled sample by a semi-automatic method based on {GMM} fitting in the color-magnitude diagram, which is then applied to other galaxies by shifting the lines to match the position of TRGB for the individual galaxy. In this way, the C-AGB and O-AGB stars with certain and typical locations are identified for all the 14 galaxies.

The C/M ratio is found to decrease with $(J-K)_0$ of TRGB, which is the indicator of metallicity, which coincides with previous studies and agrees with the expectation because C-AGB stars are more easily formed in a metal-poor environment. The quantitative relation of the C/M ratio and metallicity is derived as well. For the two large galaxies M31 and M33, the radial variation within the galaxy is presented. Within M33, this ratio increases with galactocentric radius, which agrees with the galactic chemical evolution model for spiral galaxies. Meanwhile, it is found to decrease with radius and to increase in the ring area in M31, which can be explained by the mass gradient. We suspect that stellar mass gradient may play a role, and further study is required.

\vspace{6pt}

\authorcontributions{Conceptualization, B.J.; methodology, T.R., B.J., Y.R. and M.Y.; validation, T.R., B.J., Y.R. and M.Y.; formal analysis, T.R., B.J., Y.R. and M.Y.; investigation, T.R., B.J., Y.R. and M.Y.; writing---original draft preparation, T.R.; writing---review and editing, T.R., B.J., Y.R. and M.Y.; visualization, T.R.; supervision, B.J.; project administration, B.J.; funding acquisition, B.J. All authors have read and agreed to the published version of the manuscript.}

\funding{This work is supported by NSFC 12133002, National Key R\&D Program of China No. 2019YFA0405503, and CMS-CSST-2021-A09.}

\dataavailability{This work is based on the datasets of \citetalias{ren2021red} and \mbox{\citetalias{ren2021rsglocal}} made from UKIRT and 2MASS.}

\acknowledgments{We thank Yang Chen, Jian Gao, {Yong Zhang} and  Xiaoting Fu, Haibo Yuan, Shanqin Wang, and {the anonymous referees} for their very helpful discussions and advice.}

\conflictsofinterest{The authors declare no conflict of interest.}

\begin{adjustwidth}{-\extralength}{0cm}
\printendnotes[custom] 

\reftitle{References}

\end{adjustwidth}

\begin{thebibliography}{999}


\bibitem[Herwig(2005)]{herwig2005evolution}
Herwig, F.
\newblock Evolution of asymptotic giant branch stars.
\newblock {\em Annu. Rev. Astron. Astrophys.} {\bf 2005}, {\em 43},~435--479.

\bibitem[Pinte et~al.(2009)Pinte, Harries, Min, Watson, Dullemond, Woitke,
  M{\'e}nard, and Dur{\'a}n-Rojas]{pinte2009benchmark}
Pinte, C.; Harries, T.; Min, M.; Watson, A.; Dullemond, C.; Woitke, P.;
  M{\'e}nard, F.; Dur{\'a}n-Rojas, M.
\newblock Benchmark problems for continuum radiative transfer-High optical
  depths, anisotropic scattering, and polarisation.
\newblock {\em Astron.  Astrophys.} {\bf 2009}, {\em 498},~967--980.

\bibitem[H{\"o}fner and Olofsson(2018)]{hofner2018mass}
H{\"o}fner, S.; Olofsson, H.
\newblock Mass loss of stars on the asymptotic giant branch.
\newblock {\em Astron. Astrophys. Rev.} {\bf 2018}, {\em
  26},~1--92.

\bibitem[Vassiliadis and Wood(1993)]{vassiliadis1993evolution}
Vassiliadis, E.; Wood, P.
\newblock Evolution of low-and intermediate-mass stars to the end of the
  asymptotic giant branch with mass loss.
\newblock {\em Astrophys. J.} {\bf 1993}, {\em 413},~641--657.

\bibitem[Iben and Renzini(1983)]{iben1983asymptotic}
Iben, I.; Renzini, A.
\newblock Asymptotic giant branch evolution and beyond.
\newblock {\em Annu. Rev. Astron. Astrophys.} {\bf 1983}, {\em
  21},~271--342.

\bibitem[Busso et~al.(1999)Busso, Gallino, and
  Wasserburg]{busso1999nucleosynthesis}
Busso, M.; Gallino, R.; Wasserburg, G.
\newblock Nucleosynthesis in asymptotic giant branch stars: Relevance for
  galactic enrichment and solar system formation.
\newblock {\em Annu. Rev. Astron. Astrophys.} {\bf 1999}, {\em
  37},~239--309.

\bibitem[Fishlock et~al.(2014)Fishlock, Karakas, Lugaro, and
  Yong]{fishlock2014evolution}
Fishlock, C.K.; Karakas, A.I.; Lugaro, M.; Yong, D.
\newblock Evolution and nucleosynthesis of asymptotic giant branch stellar
  models of low metallicity.
\newblock {\em Astrophys. J.} {\bf 2014}, {\em 797},~44.

\bibitem[Karakas and Lattanzio(2014)]{karakas2014dawes}
Karakas, A.I.; Lattanzio, J.C.
\newblock The Dawes review 2: Nucleosynthesis and stellar yields of low-and
  intermediate-mass single stars.
\newblock {\em Pub. Astron. Soc. Aus.} {\bf
  2014}, {\em 31}.

\bibitem[Cristallo et~al.(2015)Cristallo, Straniero, Piersanti, and
  Gobrecht]{cristallo2015evolution}
Cristallo, S.; Straniero, O.; Piersanti, L.; Gobrecht, D.
\newblock Evolution, nucleosynthesis, and yields of AGB stars at different
  metallicities. III. Intermediate-mass models, revised low-mass models, and
  the ph-FRUITY interface.
\newblock {\em Astrophys. J. Suppl. Ser.} {\bf 2015}, {\em
  219},~40.

\bibitem[Weiss and Ferguson(2009)]{weiss2009new}
Weiss, A.; Ferguson, J.W.
\newblock New asymptotic giant branch models for a range of metallicities.
\newblock {\em Astron. Astrophys.} {\bf 2009}, {\em 508},~1343--1358.

\bibitem[Trippella et~al.(2014)Trippella, Busso, Maiorca, K{\"a}ppeler,
  and Palmerini]{trippella2014s}
Trippella, O.; Busso, M.; Maiorca, E.; K{\"a}ppeler, F.; Palmerini, S.
\newblock s-Processing in AGB Stars Revisited. I. Does the Main Component
  Constrain the Neutron Source in the 13C Pocket?
\newblock {\em Astrophys. J.} {\bf 2014}, {\em 787},~41.

\bibitem[Trippella et~al.(2016)Trippella, Busso, Palmerini, Maiorca, and
  Nucci]{trippella2016s}
Trippella, O.; Busso, M.; Palmerini, S.; Maiorca, E.; Nucci, M.
\newblock s-processing in AGB stars revisited. II. Enhanced 13C production
  through MHD-induced mixing.
\newblock {\em Astrophys. J.} {\bf 2016}, {\em 818},~125.

\bibitem[Busso et~al.(2021)Busso, Vescovi, Palmerini, Cristallo, and
  Antonuccio-Delogu]{busso2021s}
Busso, M.; Vescovi, D.; Palmerini, S.; Cristallo, S.; Antonuccio-Delogu, V.
\newblock s-Processing in AGB Stars Revisited. III. Neutron captures from MHD
  mixing at different metallicities and observational constraints.
\newblock {\em Astrophys. J.} {\bf 2021}, {\em 908},~55.

\bibitem[Smolders et~al.(2012)Smolders, Neyskens, Blommaert, Hony,
  Van~Winckel, Decin, Van~Eck, Sloan, Cami, Uttenthaler,
  et~al.]{smolders2012spitzer}
Smolders, K.; Neyskens, P.; Blommaert, J.; Hony, S.; Van~Winckel, H.; Decin,
  L.; Van~Eck, S.; Sloan, G.; Cami, J.; \mbox{Uttenthaler, S.;  et~al.}
\newblock The Spitzer spectroscopic survey of S-type stars.
\newblock {\em Astron. Astrophys.} {\bf 2012}, {\em 540},~A72.

\bibitem[Cioni and Habing(2003)]{cioni2003agb}
Cioni, M.R.; Habing, H.
\newblock AGB stars in the Magellanic Clouds-I. The C/M ratio.
\newblock {\em Astron. Astrophys.} {\bf 2003}, {\em 402},~133--140.

\bibitem[Secchi(1868)]{secchi1868catalogue}
Secchi, A.
\newblock A Catalogue of Spectra of Red Stars.
\newblock {\em Mon. Not. Roy. Astron. Soc.} {\bf 1868},
  {\em 28},~196.

\bibitem[Battinelli and Demers(2005)]{battinelli2005calibration}
Battinelli, P.; Demers, S.
\newblock The calibration of the metallicity versus C/M relation.
\newblock {\em Astron. Astrophys.} {\bf 2005}, {\em 434},~657--663.

\bibitem[Cioni and Habing(2005)]{cioni2005near}
Cioni, M.R.; Habing, H.
\newblock Near-IR observations of NGC 6822: AGB stars, distance, metallicity
  and structure.
\newblock {\em Astron. Astrophys.} {\bf 2005}, {\em 429},~837--850.

\bibitem[Frogel et~al.(1990)Frogel, Mould, and
  Blanco]{frogel1990asymptotic}
Frogel, J.A.; Mould, J.; Blanco, V.
\newblock The asymptotic giant branch of Magellanic Cloud clusters.
\newblock {\em Astrophys. J.} {\bf 1990}, {\em 352},~96--122.

\bibitem[Marigo et~al.(2017)Marigo, Girardi, Bressan, Rosenfield, Aringer,
  Chen, Dussin, Nanni, Pastorelli, Rodrigues, et~al.]{marigo2017new}
Marigo, P.; Girardi, L.; Bressan, A.; Rosenfield, P.; Aringer, B.; Chen, Y.;
  Dussin, M.; Nanni, A.; Pastorelli, G.; Rodrigues, T.S.;  et~al.
\newblock A new generation of PARSEC-COLIBRI stellar isochrones including the
  TP-AGB phase.
\newblock {\em Astrophys. J.} {\bf 2017}, {\em 835},~77.

\bibitem[Palmer and Wing(1982)]{palmer1982new}
Palmer, L.; Wing, R.
\newblock A new search technique for M and C stars.
\newblock {\em Astron. J.} {\bf 1982}, {\em 87},~1739--1750.

\bibitem[Cook et~al.(1986)Cook, Aaronson, and Norris]{cook1986carbon}
Cook, K.; Aaronson, M.; Norris, J.
\newblock Carbon and M stars in nearby galaxies-A preliminary survey using a
  photometric technique.
\newblock {\em Astrophys. J.} {\bf 1986}, {\em 305},~634--644.

\bibitem[Brewer et~al.(1995)Brewer, Richer, and Crabtree]{brewer1995late}
Brewer, J.P.; Richer, H.B.; Crabtree, D.R.
\newblock Late-Type Stars in M31. I. Photometric Study of AGB Stars and
  Metallicity Gradients.
\newblock {\em Astron. J.} {\bf 1995}, {\em 109},~2480.

\bibitem[Brewer et~al.(1996)Brewer, Richer, and Crabtree]{brewer1996late}
Brewer, J.P.; Richer, H.B.; Crabtree, D.R.
\newblock Late-type stars in M31. II. C-, S-, and M-star spectra.
\newblock {\em Astron. J.} {\bf 1996}, {\em 112},~491.

\bibitem[Letarte et~al.(2002)Letarte, Demers, Battinelli, and
  Kunkel]{letarte2002extent}
Letarte, B.; Demers, S.; Battinelli, P.; Kunkel, W.
\newblock The Extent of NGC 6822 Revealed by Its C Star Population.
\newblock {\em Astron. J.} {\bf 2002}, {\em 123},~832.

\bibitem[Groenewegen(2006)]{groenewegen2006agb}
Groenewegen, M.
\newblock AGB stars in the Local Group, and beyond. In {\em Planetary Nebulae
  Beyond the Milky Way}; Springer:  Berlin/Heidelberg, Germany, 
 2006; pp. 108--120.

\bibitem[Demers et~al.(2006)Demers, Battinelli, and
  Artigau]{demers2006carbon}
Demers, S.; Battinelli, P.; Artigau, E.
\newblock Carbon stars in the outer spheroid of NGC 6822.
\newblock {\em Astron. Astrophys.} {\bf 2006}, {\em 456},~905--910.

\bibitem[Davidge(1998)]{davidge1998evolved}
Davidge, T.
\newblock The evolved red stellar contents of the Sculptor group galaxies NGC
  55, NGC 300, and NGC 7793.
\newblock {\em Astrophys. J.} {\bf 1998}, {\em 497},~650.

\bibitem[Davidge(2000)]{davidge2000evolved}
Davidge, T.
\newblock The evolved red stellar content of M32.
\newblock {\em Pub. Astrono. Soc. Pac.} {\bf
  2000}, {\em 112},~1177.

\bibitem[Davidge(2003)]{davidge2003asymptotic}
Davidge, T.J.
\newblock The Asymptotic Giant Branch of NGC 205: The Characteristics of Carbon
  Stars and M Giants Identified from JHK'Images.
\newblock {\em Astrophys. J.} {\bf 2003}, {\em 597},~289.

\bibitem[Davidge and Rigaut(2004)]{davidge2004photometric}
Davidge, T.; Rigaut, F.
\newblock Photometric variability among the brightest asymptotic giant branch
  stars near the center of M32.
\newblock {\em Astrophys. J.} {\bf 2004}, {\em 607},~L25.

\bibitem[Davidge(2005)]{davidge2005disk}
Davidge, T.
\newblock The disk and extraplanar regions of NGC 55.
\newblock {\em Astrophys. J.} {\bf 2005}, {\em 622},~279.

\bibitem[Cioni et~al.(2008)Cioni, Irwin, Ferguson, McConnachie, Conn,
  Huxor, Ibata, Lewis, and Tanvir]{cioni2008agb}
Cioni, M.R.; Irwin, M.; Ferguson, A.; McConnachie, A.; Conn, B.; Huxor, A.;
  Ibata, R.; Lewis, G.; Tanvir, N.
\newblock AGB stars as tracers of metallicity and mean age across M 33.
\newblock {\em Astron. Astrophys.} {\bf 2008}, {\em 487},~131--146.

\bibitem[Sibbons et~al.(2012)Sibbons, Ryan, Cioni, Irwin, and
  Napiwotzki]{sibbons2012agb}
Sibbons, L.; Ryan, S.G.; Cioni, M.R.; Irwin, M.; Napiwotzki, R.
\newblock The AGB population of NGC 6822: distribution and the C/M ratio from
  JHK photometry.
\newblock {\em Astron. Astrophys.} {\bf 2012}, {\em 540},~A135.

\bibitem[Sibbons et~al.(2015)Sibbons, Ryan, Irwin, and
  Napiwotzki]{sibbons2015agb}
Sibbons, L.; Ryan, S.G.; Irwin, M.; Napiwotzki, R.
\newblock The AGB population in IC 1613 using JHK photometry.
\newblock {\em Astron. Astrophys.} {\bf 2015}, {\em 573},~A84.

\bibitem[Battinelli et~al.(2007)Battinelli, Demers, and
  Mannucci]{battinelli2007assessment}
Battinelli, P.; Demers, S.; Mannucci, F.
\newblock The assessment of the near infrared identification of carbon stars-I.
  The Local Group galaxies WLM, IC 10 and NGC 6822.
\newblock {\em Astron. Astrophys.} {\bf 2007}, {\em 474},~35--41.

\bibitem[Pastorelli et~al.(2019)Pastorelli, Marigo, Girardi, Chen, Rubele,
  Trabucchi, Aringer, Bladh, Bressan, Montalb{\'a}n,
  et~al.]{pastorelli2019constraining}
Pastorelli, G.; Marigo, P.; Girardi, L.; Chen, Y.; Rubele, S.; Trabucchi, M.;
  Aringer, B.; Bladh, S.; Bressan, A.; Montalb{\'a}n, J.;  et~al.
\newblock Constraining the thermally pulsing asymptotic giant branch phase with
  resolved stellar populations in the Small Magellanic Cloud.
\newblock {\em Mon. Not. Roy. Astron. Soc.} {\bf 2019},
  {\em 485},~5666--5692.

\bibitem[Pastorelli et~al.(2020)Pastorelli, Marigo, Girardi, Aringer,
  Chen, Rubele, Trabucchi, Bladh, Boyer, Bressan,
  et~al.]{pastorelli2020constraining}
Pastorelli, G.; Marigo, P.; Girardi, L.; Aringer, B.; Chen, Y.; Rubele, S.;
  Trabucchi, M.; Bladh, S.; Boyer, M.L.; Bressan, A.;  et~al.
\newblock Constraining the thermally pulsing asymptotic giant branch phase with
  resolved stellar populations in the Large Magellanic Cloud.
\newblock {\em Mon. Not. Roy. Astron. Soc.} {\bf 2020},
  {\em 498},~3283--3301.

\bibitem[Ren et~al.(2021{\natexlab{a}})Ren, Jiang, Yang, Wang, Jian, and
  Ren]{ren2021red}
Ren, Y.; Jiang, B.; Yang, M.; Wang, T.; Jian, M.; Ren, T.
\newblock Red Supergiants in M31 and M33. I. The Complete Sample.
\newblock {\em Astrophys. J.} {\bf 2021}, {\em 907},~18.

\bibitem[Ren et~al.(2021{\natexlab{b}})Ren, Jiang, Yang, Wang, and
  Ren]{ren2021rsglocal}
Ren, Y.; Jiang, B.; Yang, M.; Wang, T.; Ren, T.
\newblock The Sample of Red Supergiants in 12 
  Group.
\newblock {\em The Astrophys. J.} {\bf 2021}, {\em 923},~232.

\bibitem[Allard and Hauschildt(1995)]{allard1996model}
Allard, F.; Hauschildt, P.H.
\newblock Model Atmospheres for M (Sub) Dwarfs: I. The base model grid.
\newblock {\em Astrophys. J.} {\bf 1995}, {\em 445},~433.

\bibitem[Bessell and Brett(1988)]{bessell1988jhklm}
Bessell, M.; Brett, J.
\newblock JHKLM photometry: standard systems, passbands, and intrinsic colors.
\newblock {\em Pub. Astron. Soc. Pac.} {\bf
  1988}, {\em 100},~1134.

\bibitem[Schlegel et~al.(1998)Schlegel, Finkbeiner, and
  Davis]{schlegel1998maps}
Schlegel, D.J.; Finkbeiner, D.P.; Davis, M.
\newblock Maps of dust infrared emission for use in estimation of reddening and
  cosmic microwave background radiation foregrounds.
\newblock {\em Astrophys. J.} {\bf 1998}, {\em 500},~525.

\bibitem[Green et~al.(2019)Green, Schlafly, Zucker, Speagle, and
  Finkbeiner]{green20193d}
Green, G.M.; Schlafly, E.; Zucker, C.; Speagle, J.S.; Finkbeiner, D.
\newblock A 3D Dust Map Based on Gaia, Pan-STARRS 1, and 2MASS.
\newblock {\em Astrophys. J.} {\bf 2019}, {\em 887},~93.

\bibitem[Skowron et~al.(2021)Skowron, Skowron, Udalski, Szyma{\'n}ski,
  Soszy{\'n}ski, Ulaczyk, Poleski, Koz{\l}owski, Pietrukowicz, Mr{\'o}z,
  et~al.]{skowron2021ogle}
Skowron, D.; Skowron, J.; Udalski, A.; Szyma{\'n}ski, M.; Soszy{\'n}ski, I.;
  Ulaczyk, K.; Poleski, R.; Koz{\l}owski, S.; Pietrukowicz, P.; Mr{\'o}z, P.;
  et~al.
\newblock OGLE-ing the Magellanic System: Optical Reddening Maps of the Large
  and Small Magellanic Clouds from Red Clump Stars.
\newblock {\em Astrophys. J. Suppl. Ser.} {\bf 2021}, {\em
  252},~23.

\bibitem[Bellazzini et~al.(2004)Bellazzini, Ferraro, Sollima, Pancino, and
  Origlia]{bellazzini2004calibration}
Bellazzini, M.; Ferraro, F.; Sollima, A.; Pancino, E.; Origlia, L.
\newblock The calibration of the RGB Tip as a Standard Candle-Extension to Near
  Infrared colors and higher metallicity.
\newblock {\em Astron. Astrophys.} {\bf 2004}, {\em 424},~199--211.

\bibitem[Pietrzy{\'n}ski et~al.(2019)Pietrzy{\'n}ski, Graczyk, Gallenne,
  Gieren, Thompson, Pilecki, Karczmarek, G{\'o}rski, Suchomska, Taormina,
  et~al.]{pietrzynski2019distance}
Pietrzy{\'n}ski, G.; Graczyk, D.; Gallenne, A.; Gieren, W.; Thompson, I.;
  Pilecki, B.; Karczmarek, P.; G{\'o}rski, M.; Suchomska, K.; Taormina, M.;
  et~al.
\newblock A distance to the Large Magellanic Cloud that is precise to one per
  cent.
\newblock {\em Nature} {\bf 2019}, {\em 567},~200--203.

\bibitem[Graczyk et~al.(2020)Graczyk, Pietrzy{\'n}ski, Thompson, Gieren,
  Zgirski, Villanova, G{\'o}rski, Wielg{\'o}rski, Karczmarek, Narloch,
  et~al.]{graczyk2020distance}
Graczyk, D.; Pietrzy{\'n}ski, G.; Thompson, I.B.; Gieren, W.; Zgirski, B.;
  Villanova, S.; G{\'o}rski, M.; Wielg{\'o}rski, P.; Karczmarek, P.; Narloch,
  W.;  et~al.
\newblock A Distance Determination to the Small Magellanic Cloud with an
  Accuracy of Better than Two Percent Based on Late-type Eclipsing Binary
  Stars.
\newblock {\em Astrophys. J.} {\bf 2020}, {\em 904},~13.

\bibitem[Skrutskie et~al.(2006)Skrutskie, Cutri, Stiening, Weinberg,
  Schneider, Carpenter, Beichman, Capps, Chester, Elias,
  et~al.]{skrutskie2006two}
Skrutskie, M.; Cutri, R.; Stiening, R.; Weinberg, M.; Schneider, S.; Carpenter,
  J.; Beichman, C.; Capps, R.; Chester, T.; Elias, J.;  et~al.
\newblock The two micron all sky survey (2MASS).
\newblock {\em Astron. J.} {\bf 2006}, {\em 131},~1163.

\bibitem[Hodgkin et~al.(2009)Hodgkin, Irwin, Hewett, and
  Warren]{hodgkin2009ukirt}
Hodgkin, S.; Irwin, M.; Hewett, P.; Warren, S.
\newblock The UKIRT wide field camera ZYJHK photometric system: calibration
  from 2MASS.
\newblock {\em Mon. Not. Roy. Astron. Soc.} {\bf 2009},
  {\em 394},~675--692.

\bibitem[Wang et~al.(2021)Wang, Jiang, Ren, Yang, and Li]{wang2021red}
Wang, T.; Jiang, B.; Ren, Y.; Yang, M.; Li, J.
\newblock Red Supergiants in M31 and M33. II. The Mass-loss Rate.
\newblock {\em Astrophys. J.} {\bf 2021}, {\em 912},~112.

\bibitem[Hirschauer et~al.(2020)Hirschauer, Gray, Meixner, Jones,
  Srinivasan, Boyer, and Sargent]{hirschauer2020dusty}
Hirschauer, A.S.; Gray, L.; Meixner, M.; Jones, O.C.; Srinivasan, S.; Boyer,
  M.L.; Sargent, B.
\newblock Dusty Stellar Birth and Death in the Metal-poor Galaxy NGC 6822.
\newblock {\em Astrophys. J.} {\bf 2020}, {\em 892},~91.

\bibitem[Rosenfield et~al.(2016)Rosenfield, Marigo, Girardi, Dalcanton,
  Bressan, Williams, and Dolphin]{rosenfield2016evolution}
Rosenfield, P.; Marigo, P.; Girardi, L.; Dalcanton, J.J.; Bressan, A.;
  Williams, B.F.; Dolphin, A.
\newblock Evolution of thermally pulsing asymptotic giant branch stars. V.
  Constraining the mass loss and lifetimes of intermediate-mass,
  low-metallicity AGB stars.
\newblock {\em Astrophys. J.} {\bf 2016}, {\em 822},~73.

\bibitem[Fraley and Raftery(2002)]{fraley2002model}
Fraley, C.; Raftery, A.E.
\newblock Model-based clustering, discriminant analysis, and density
  estimation.
\newblock {\em J. Am. Stat. Assoc.} {\bf 2002},
  {\em 97},~611--631.

\bibitem[Marigo et~al.(2003)Marigo, Girardi, and Chiosi]{marigo2003red}
Marigo, P.; Girardi, L.; Chiosi, C.
\newblock The red tail of carbon stars in the LMC: Models meet 2MASS and DENIS
  observations.
\newblock {\em Astron. Astrophys.} {\bf 2003}, {\em 403},~225--237.

\bibitem[Da~Costa and Armandroff(1990)]{da1990standard}
Da~Costa, G.; Armandroff, T.
\newblock Standard globular cluster giant branches in the (MI,/VI/sub O) plane.
\newblock {\em Astron. J.} {\bf 1990}, {\em 100},~162--181.

\bibitem[Lee et~al.(1993)Lee, Freedman, and Madore]{lee1993tip}
Lee, M.G.; Freedman, W.L.; Madore, B.F.
\newblock The tip of the red giant branch as a distance indicator for resolved
  galaxies.
\newblock {\em Astrophys. J.} {\bf 1993}, {\em 417},~553.

\bibitem[Ivanov et~al.(2000)Ivanov, Borissova, Alonso-Herrero, and
  Russeva]{ivanov2000extending}
Ivanov, V.D.; Borissova, J.; Alonso-Herrero, A.; Russeva, T.
\newblock Extending the Red Giant Branch versus Metallicity Calibration toward
  Metal-poor Systems: Near-Infrared Photometry of the Galactic Globular
  Clusters M56 and M15.
\newblock {\em Astron. J.} {\bf 2000}, {\em 119},~2274.

\bibitem[Cioni(2009)]{cioni2009metallicity}
Cioni, M.R.
\newblock The metallicity gradient as a tracer of history and structure: the
  Magellanic Clouds and M33 galaxies.
\newblock {\em Astron. Astrophys.} {\bf 2009}, {\em 506},~1137--1146.

\bibitem[Kang et~al.(2006)Kang, Sohn, Kim, Rhee, Kim, Hwang, Lee, Kim, and
  Chun]{kang2006evolved}
Kang, A.; Sohn, Y.J.; Kim, H.I.; Rhee, J.; Kim, J.W.; Hwang, N.; Lee, M.; Kim,
  Y.C.; Chun, M.S.
\newblock The evolved asymptotic giant branch stars in the central bar of the
  dwarf irregular galaxy NGC 6822.
\newblock {\em Astron. Astrophys.} {\bf 2006}, {\em 454},~717--727.

\bibitem[Kang et~al.(2005)Kang, Sohn, Rhee, Shin, Chun, and
  Kim]{kang2005near}
Kang, A.; Sohn, Y.J.; Rhee, J.; Shin, M.; Chun, M.S.; Kim, H.I.
\newblock Near-IR photometry of asymptotic giant branch stars in the dwarf
  elliptical galaxy NGC 185.
\newblock {\em Astron. Astrophys.} {\bf 2005}, {\em 437},~61--68.

\bibitem[Boyer et~al.(2017)Boyer, McQuinn, Groenewegen, Zijlstra,
  Whitelock, Van~Loon, Sonneborn, Sloan, Skillman, Meixner,
  et~al.]{boyer2017infrared}
Boyer, M.; McQuinn, K.; Groenewegen, M.; Zijlstra, A.; Whitelock, P.; Van~Loon,
  J.T.; Sonneborn, G.; Sloan, G.; Skillman, E.; Meixner, M.;  et~al.
\newblock An Infrared Census of DUST in Nearby Galaxies with Spitzer
  (DUSTiNGS). IV. Discovery of High-redshift AGB Analogs.
\newblock {\em Astrophys. J.} {\bf 2017}, {\em 851},~152.

\bibitem[Battinelli et~al.(2003)Battinelli, Demers, and
  Letarte]{battinelli2003carbon}
Battinelli, P.; Demers, S.; Letarte, B.
\newblock Carbon star survey in the Local Group. V. The outer disk of M31.
\newblock {\em Astron. J.} {\bf 2003}, {\em 125},~1298.

\bibitem[Jones et~al.(2018)Jones, Maclay, Boyer, Meixner, McDonald, and
  Meskhidze]{jones2018near}
Jones, O.C.; Maclay, M.T.; Boyer, M.L.; Meixner, M.; McDonald, I.; Meskhidze,
  H.
\newblock Near-Infrared Stellar Populations in the metal-poor, Dwarf irregular
  Galaxies Sextans A and Leo A.
\newblock {\em Astrophys. J.} {\bf 2018}, {\em 854},~117.

\bibitem[Valcheva et~al.(2007)Valcheva, Ivanov, Ovcharov, and
  Nedialkov]{valcheva2007carbon}
Valcheva, A.; Ivanov, V.; Ovcharov, E.; Nedialkov, P.
\newblock Carbon stars and C/M ratio in the WLM dwarf irregular galaxy.
\newblock {\em Astron. Astrophys.} {\bf 2007}, {\em 466},~501--507.

\bibitem[Freedman et~al.(2020)Freedman, Madore, Hoyt, Jang, Beaton, Lee,
  Monson, Neeley, and Rich]{freedman2020calibration}
Freedman, W.L.; Madore, B.F.; Hoyt, T.; Jang, I.S.; Beaton, R.; Lee, M.G.;
  Monson, A.; Neeley, J.; Rich, J.
\newblock Calibration of the Tip of the red giant branch.
\newblock {\em Astrophys. J.} {\bf 2020}, {\em 891},~57.

\bibitem[Cioni et~al.(2006)Cioni, Girardi, Marigo, and
  Habing]{cioni2006agb}
Cioni, M.R.; Girardi, L.; Marigo, P.; Habing, H.
\newblock AGB stars in the Magellanic Clouds-II. The rate of star formation
  across the LMC.
\newblock {\em Astron. Astrophys.} {\bf 2006}, {\em 448},~77--91.

\bibitem[Wang and Chen(2019)]{wang2019optical}
Wang, S.; Chen, X.
\newblock The optical to mid-infrared extinction law based on the APOGEE, Gaia
  DR2, Pan-STARRS1, SDSS, APASS, 2MASS, and WISE surveys.
\newblock {\em Astrophys. J.} {\bf 2019}, {\em 877},~116.

\bibitem[Bressan et~al.(2012)Bressan, Marigo, Girardi, Salasnich,
  Dal~Cero, Rubele, and Nanni]{bressan2012parsec}
Bressan, A.; Marigo, P.; Girardi, L.; Salasnich, B.; Dal~Cero, C.; Rubele, S.;
  Nanni, A.
\newblock PARSEC: stellar tracks and isochrones with the PAdova and TRieste
  Stellar Evolution Code.
\newblock {\em Mon. Not. Roy. Astron. Soc.} {\bf 2012},
  {\em 427},~127--145.

\bibitem[Marigo et~al.(2013)Marigo, Bressan, Nanni, Girardi, and
  Pumo]{marigo2013evolution}
Marigo, P.; Bressan, A.; Nanni, A.; Girardi, L.; Pumo, M.L.
\newblock Evolution of thermally pulsing asymptotic giant branch stars--I. The
  colibri code.
\newblock {\em Mon. Not. Roy. Astron. Soc.} {\bf 2013},
  {\em 434},~488--526.

\bibitem[Tang et~al.(2014)Tang, Bressan, Rosenfield, Slemer, Marigo,
  Girardi, and Bianchi]{tang2014new}
Tang, J.; Bressan, A.; Rosenfield, P.; Slemer, A.; Marigo, P.; Girardi, L.;
  Bianchi, L.
\newblock New PARSEC evolutionary tracks of massive stars at low metallicity:
  testing canonical stellar evolution in nearby star-forming dwarf galaxies.
\newblock {\em Mon. Not. Roy. Astron. Soc.} {\bf 2014},
  {\em 445},~4287--4305.

\bibitem[Chen et~al.(2014)Chen, Girardi, Bressan, Marigo, Barbieri, and
  Kong]{chen2014improving}
Chen, Y.; Girardi, L.; Bressan, A.; Marigo, P.; Barbieri, M.; Kong, X.
\newblock Improving PARSEC models for very low mass stars.
\newblock {\em Mon. Not. Roy. Astron. Soc.} {\bf 2014},
  {\em 444},~2525--2543.

\bibitem[Chen et~al.(2015)Chen, Bressan, Girardi, Marigo, Kong, and
  Lanza]{chen2015parsec}
Chen, Y.; Bressan, A.; Girardi, L.; Marigo, P.; Kong, X.; Lanza, A.
\newblock PARSEC evolutionary tracks of massive stars up to 350 M$_\odot$ at
  metallicities 0.0001 $\leq$ Z $\leq$ 0.04.
\newblock {\em Mon. Not. Roy. Astron. Soc.} {\bf 2015},
  {\em 452},~1068--1080.

\bibitem[Kroupa(2001)]{kroupa2001variation}
Kroupa, P.
\newblock On the variation of the initial mass function.
\newblock {\em Mon. Not. Roy. Astron. Soc.} {\bf 2001},
  {\em 322},~231--246.

\bibitem[Kroupa(2002)]{kroupa2002initial}
Kroupa, P.
\newblock The initial mass function of stars: evidence for uniformity in
  variable systems.
\newblock {\em Science} {\bf 2002}, {\em 295},~82--91.

\bibitem[Groenewegen(2006)]{groenewegen2006mid}
Groenewegen, M.
\newblock The mid-and far-infrared colours of AGB and post-AGB stars.
\newblock {\em Astron. Astrophys.} {\bf 2006}, {\em 448},~181--187.

\bibitem[Blanco et~al.(1980)Blanco, McCarthy, and
  Blanco]{blanco1980carbon}
Blanco, V.; McCarthy, M.; Blanco, B.
\newblock Carbon and late M-type stars in the Magellanic Clouds.
\newblock {\em Astrophys. J.} {\bf 1980}, {\em 242},~938--964.

\bibitem[Kacharov et~al.(2012)Kacharov, Rejkuba, and
  Cioni]{kacharov2012spectra}
Kacharov, N.; Rejkuba, M.; Cioni, M.R.
\newblock Spectra probing the number ratio of C-to M-type AGB stars in the NGC
  6822 galaxy.
\newblock {\em Astron. Astrophys.} {\bf 2012}, {\em 537},~A108.

\bibitem[Blum et~al.(2006)Blum, Mould, Olsen, Frogel, Werner, Meixner,
  Markwick-Kemper, Indebetouw, Whitney, Meade, et~al.]{blum2006spitzer}
Blum, R.; Mould, J.; Olsen, K.; Frogel, J.; Werner, M.; Meixner, M.;
  Markwick-Kemper, F.; Indebetouw, R.; Whitney, B.; \mbox{Meade, M.;  et~al}.
\newblock Spitzer SAGE Survey of the Large Magellanic Cloud. II. Evolved Stars
  and Infrared Color-Magnitude Diagrams.
\newblock {\em Astron. J.} {\bf 2006}, {\em 132},~2034.

\bibitem[Boyer et~al.(2011)Boyer, Srinivasan, Van~Loon, McDonald, Meixner,
  Zaritsky, Gordon, Kemper, Babler, Block, et~al.]{boyer2011surveying}
Boyer, M.L.; Srinivasan, S.; Van~Loon, J.T.; McDonald, I.; Meixner, M.;
  Zaritsky, D.; Gordon, K.D.; Kemper, F.; Babler, B.; \mbox{Block, M.;  et~al.}
\newblock Surveying the agents of galaxy evolution in the tidally stripped, low
  metallicity Small Magellanic Cloud (SAGE-SMC). II. Cool evolved stars.
\newblock {\em Astron. J.} {\bf 2011}, {\em 142},~103.

\bibitem[Dell'Agli et~al.(2015)Dell'Agli, Garc{\'\i}a-Hern{\'a}ndez,
  Ventura, Schneider, Di~Criscienzo, and Rossi]{dell2015agb}
Dell'Agli, F.; Garc{\'\i}a-Hern{\'a}ndez, D.; Ventura, P.; Schneider, R.;
  Di~Criscienzo, M.; Rossi, C.
\newblock AGB stars in the SMC: evolution and dust properties based on Spitzer
  observations.
\newblock {\em Mon. Not. Roy. Astron. Soc.} {\bf 2015},
  {\em 454},~4235--4249.

\bibitem[Serenelli et~al.(2017)Serenelli, Weiss, Cassisi, Salaris, and
  Pietrinferni]{serenelli2017brightness}
Serenelli, A.; Weiss, A.; Cassisi, S.; Salaris, M.; Pietrinferni, A.
\newblock The brightness of the red giant branch tip-Theoretical framework, a
  set of reference models, and predicted observables.
\newblock {\em Astron. Astrophys.} {\bf 2017}, {\em 606},~A33.

\bibitem[Hoyt et~al.(2018)Hoyt, Freedman, Madore, Seibert, Beaton, Hatt,
  Jang, Lee, Monson, and Rich]{hoyt2018near}
Hoyt, T.J.; Freedman, W.L.; Madore, B.F.; Seibert, M.; Beaton, R.L.; Hatt, D.;
  Jang, I.S.; Lee, M.G.; Monson, A.J.; Rich, J.A.
\newblock The near-infrared tip of the red giant branch. II. An absolute
  calibration in the Large Magellanic Cloud.
\newblock {\em Astrophys. J.} {\bf 2018}, {\em 858},~12.

\bibitem[Marigo et~al.(2008)Marigo, Girardi, Bressan, Groenewegen, Silva,
  and Granato]{marigo2008evolution}
Marigo, P.; Girardi, L.; Bressan, A.; Groenewegen, M.A.; Silva, L.; Granato,
  G.L.
\newblock Evolution of asymptotic giant branch stars-II. Optical to
  far-infrared isochrones with improved tp-agb models.
\newblock {\em Astron. Astrophys.} {\bf 2008}, {\em 482},~883--905.

\bibitem[Boyer et~al.(2015)Boyer, McDonald, Srinivasan, Zijlstra, van
  Loon, Olsen, and Sonneborn]{boyer2015identification}
Boyer, M.L.; McDonald, I.; Srinivasan, S.; Zijlstra, A.; van Loon, J.T.; Olsen,
  K.A.; Sonneborn, G.
\newblock Identification of a class of low-mass asymptotic giant branch stars
  struggling to become carbon stars in the magellanic clouds.
\newblock {\em Astrophys. J.} {\bf 2015}, {\em 810},~116.

\bibitem[Hamren et~al.(2015)Hamren, Rockosi, Guhathakurta, Boyer, Smith,
  Dalcanton, Gregersen, Seth, Lewis, Williams, et~al.]{hamren2015spectroscopic}
Hamren, K.M.; Rockosi, C.M.; Guhathakurta, P.; Boyer, M.L.; Smith, G.H.;
  Dalcanton, J.J.; Gregersen, D.; Seth, A.C.; Lewis, A.R.; Williams, B.F.;
  et~al.
\newblock a Spectroscopic and Photometric Exploration of the C/M Ratio in the
  Disk of M31.
\newblock {\em Astrophys. J.} {\bf 2015}, {\em 810},~60.

\bibitem[Pena and Flores-Dur{\'a}n(2019)]{pena2019metallicity}
Pena, M.; Flores-Dur{\'a}n, S.N.
\newblock Metallicity gradients in M31, M 33, NGC 300 and the milky way using
  abundances of different elements.
\newblock {\em Revista Mexicana de Astronom{\'\i}a y Astrof{\'\i}sica} {\bf
  2019}, {\em 55},~255--271.

\bibitem[Feast et~al.(2010)Feast, Abedigamba, and
  Whitelock]{feast2010there}
Feast, M.W.; Abedigamba, O.P.; Whitelock, P.A.
\newblock Is there a metallicity gradient in the Large Magellanic Cloud?
\newblock {\em Mon. Not. Roy. Astron. Soc. Lett.} {\bf
  2010}, {\em 408},~L76--L79.

\end{thebibliography}
\end{document}